\documentclass[lettersize]{emulateapj}

\usepackage{natbib}
\usepackage{amsmath}
\usepackage{multirow}
\usepackage{float}
\usepackage{color}
\usepackage{afterpage}
\usepackage[abs]{overpic}
\usepackage{CJK}

\def\be{\begin{equation}}
\def\ee{\end{equation}}
\def\bea{\begin{eqnarray}}
\def\eea{\end{eqnarray}}



\def \logTd6 {\hbox{log$( T/6 \kev)$} }

\def\myputfigure#1#2#3#4#5%
{\vskip#5pt\makebox[0pt]{\hskip#2in
\includegraphics[width=#3\textwidth]{#1}}\vskip#4pt\hfill}

\def \arcmin     { ^{\prime} }
\def \arcsec    {^{\prime\prime}}

\def \etal      {et~al.\ }

\def \kev       {{\rm\ keV}}

\definecolor{Red}{rgb}{1,0,0}
\definecolor{Blue}{rgb}{0,0,1}
\definecolor{Green}{rgb}{0,1,0}
\definecolor{magenta}{rgb}{1,0,.6}
\definecolor{lightblue}{rgb}{0,.5,1}
\definecolor{lightpurple}{rgb}{.6,.4,1}
\definecolor{gold}{rgb}{.6,.5,0}
\definecolor{orange}{rgb}{1,0.4,0}
\definecolor{hotpink}{rgb}{1,0,0.5}
\definecolor{newcolor2}{rgb}{.5,.3,.5}
\definecolor{newcolor}{rgb}{0,.3,1}
\definecolor{newcolor3}{rgb}{1,0,.35}
\definecolor{darkgreen1}{rgb}{0, .35, 0}
\definecolor{darkgreen}{rgb}{0, .6, 0}
\definecolor{darkred}{rgb}{.75,0,0}

\newcommand{\mpc}{\ensuremath{\, h^{-1}\,\mathrm{Mpc} }}
\newcommand{\mpccube}{\ensuremath{\, h^{-3}\,\mathrm{Mpc}^3 }}
\newcommand{\kpc}{\, h^{-1}\,\mathrm{kpc} }
\newcommand{\kms}{\, \mathrm{km \; s^{-1}}}

\newcommand{\snr}{\ensuremath{\mathrm{S/N}}}
\newcommand{\lya}{Ly$\alpha$}

\newcommand{\beq}{\begin{equation}}
\newcommand{\eeq}{\end{equation}}
\newcommand{\bc}{\begin{center}}
\newcommand{\ec}{\end{center}}
\newcommand{\bfig}{\begin{figure}}
\newcommand{\efig}{\end{figure}}

\newcommand{\lambrest}{\ensuremath{\lambda_\mathrm{rest}}}

\newcommand{\lambalpha}{\ensuremath{\lambda_{\alpha}}}

\newcommand{\zbg}{\ensuremath{z_\mathrm{bg}}}

\newcommand{\ang}{\ensuremath{\mathrm{\AA}}}

\newcommand{\nlos}{\ensuremath{n_\mathrm{los}}}
\newcommand{\mlim}{\ensuremath{m_\mathrm{lim}}}
\newcommand{\glim}{\ensuremath{g_\mathrm{lim}}}
\newcommand{\dperp}{\ensuremath{\langle d_{\perp} \rangle}}
\newcommand{\sigthreed}{\ensuremath{\epsilon_{\rm 3D}}}

\newcommand{\texp}{\ensuremath{t_\mathrm{exp}}}
\newcommand{\delorig}{\ensuremath{\delta_F^{\rm orig}}}
\newcommand{\delrecon}{\ensuremath{\delta_F^{\rm rec}}}
\newcommand{\snreps}{\ensuremath{\mathrm{SNR}_{\epsilon}}}

\newcommand{\hrs}{\mathrm{hrs}}
\newcommand{\dm}{\mathrm{dm}}

\newcommand{\persqdeg}{\mathrm{deg}^{-2}}
\newcommand{\sqdeg}{\mathrm{deg}^{2}}
\newcommand{\cmd}{\mathbf{C}_\mathrm{MD}}
\newcommand{\cdd}{\mathbf{C}_\mathrm{DD}}

\def \etal {et~al.}

\slugcomment{Accepted by ApJ}
\shorttitle{Requirements for \lya\ Forest Tomography}
\shortauthors{Lee \etal}

\begin{document}

\title{Observational Requirements for Lyman-$\alpha$ Forest Tomographic Mapping \\ of Large-Scale Structure at \lowercase{z} $\sim 2$ }
\author{Khee-Gan Lee\altaffilmark{1}, 
Joseph F. Hennawi\altaffilmark{1}, 
Martin White\altaffilmark{2,3}, 
Rupert A.C.~Croft\altaffilmark{4}, Melih Ozbek\altaffilmark{4}}
\altaffiltext{1}{Max Planck Institute for Astronomy, K\"{o}nigstuhl 17, D-69117 Heidelberg, Germany}
\altaffiltext{2}{E.O. Lawrence Berkeley National Lab, 1 Cyclotron Rd., Berkeley, CA, 94720, USA}
\altaffiltext{3}{Department of Astronomy, University of California at Berkeley, B-20 Hearst Field Annex \# 3411,
Berkeley, CA 94720}
\altaffiltext{4}{Department of Physics, Carnegie-Mellon University, 5000 Forbes Avenue, Pittsburgh, PA 15213}
\email{lee@mpia.de}

\begin{abstract}

The $z \gtrsim 2$ \lya\ forest traces the underlying dark-matter
distribution on large scales and, given sufficient sightlines, can
be used to create 3D maps of large-scale structure.
We examine the observational requirements to construct such maps
and estimate the signal-to-noise as a function of exposure time and
sightline density.
Sightline densities at $z = 2.25$ are $\nlos
\approx [360, 1200,3300]\,\persqdeg$ at limiting magnitudes of $g
=[24.0, 24.5,25.0]$, resulting in transverse sightline separations of
$\dperp \approx [3.6, 1.9, 1.2]\,\mpc$, which roughly sets the
reconstruction scale. We simulate these reconstructions using mock
spectra with realistic noise properties, and find that
spectra with $\mathrm{S/N} \approx 4$ per angstrom can be used to
generate maps that clearly trace the underlying dark-matter at
overdensities of $\rho/\langle \rho \rangle \sim 1$.
For the VLT/VIMOS spectrograph, exposure times
$\texp = [4, 6, 10]\,\hrs$ are sufficient for maps with
spatial resolution $\sigthreed = [5.0, 3.2, 2.3]\,\mpc$.  Assuming
$\sim 250\,\mpc$ is probed along the line-of-sight,
$ 1\,\mathrm{deg}^2$ of survey area would cover a comoving volume
of $\approx 10^6\,\mpccube$ at $\langle z \rangle \sim 2.3$,
enabling efficient mapping of large volumes with 8-10m telescopes.
These maps could be
used to study galaxy environments, detect proto-clusters, and study
the topology of large-scale structure at high-z.

\end{abstract}

\keywords{ cosmology:observations --- galaxies:high-redshift --- intergalactic medium --- 
quasars: absorption lines --- surveys --- techniques:spectroscopic }

\section{Introduction}

Cosmography -- the cartography of the Universe -- has been a primary goal of astronomical surveys
ever since the first telescopic attempts to map the structure of the Milky Way \citep{herschel:1785}. 
These efforts were naturally limited to our galaxy until the nearly contemporaneous discoveries of the extragalactic 
Universe and cosmological redshifts \citep{hubble:1926, hubble:1929}, which expanded the milieu of interest towards the large-scale structure of 
the galaxy distribution.
However, the cartography of 3D large-scale structure beyond the $d \lesssim 10 \,\mathrm{Mpc}$ Local Universe 
remained challenging due to the 
inefficiency of obtaining galaxy redshifts with photographic plates: for example, the 920 redshifts listed in \citet{humason:1956} 
represented
a heroic effort, spanning two decades, to obtain $5-10$hr exposures on $V \sim 12$ galaxies. 
The advent of photoelectric detectors in astronomical spectrographs precipitated a quantum leap in cosmography
starting in the early 1980s,
with progressively larger galaxy redshift surveys ranging from CfA Redshift Survey \citep{davis:1982,geller:1989}, Las Campanas Redshift Survey \citep{shectman:1996}, 
through to the 2dFGRS \citep{colless:2001}, SDSS-I and -II \citep{abazajian:2009}, and GAMA \citep{driver:2011}
surveys that have now comprehensively mapped out the 
galaxy distribution out to $z\sim 0.3$.
The resulting 3D galaxy distributions from these surveys have revealed the beautiful filamentary `cosmic web' 
structure agreeing with theoretical predictions for a cold dark matter (CDM) universe, and forms one of the pillars of the current
cosmological paradigm. 
Apart from the statistical clustering measurements \citep[e.g.,][]{tegmark:2004,cole:2005,reid:2012}, the maps from 
galaxy redshift surveys have enabled the study of galaxy properties in the context of large-scale environment
\citep[e.g.,][]{lewis:2002,gomez:2003,kauffmann:2004,blanton:2005}.

However, since galaxy surface brightness scales with redshift as
$\propto (1+z)^{-4}$, it becomes increasingly expensive to obtain
redshifts for higher-redshift galaxies.  Various redshift surveys are
underway to measure samples of tracer galaxies \citep[e.g. luminous
  red galaxies or emission-line galaxies, see][]{hill:2008,
  drinkwater:2010, ahn:2012, comparat:2013} to measure large-scale
($\gtrsim 20\,\mpc$) clustering out to $z \sim 1$, but it is
considerably more difficult to obtain volume-limited samples of
galaxies at sufficient space densities to directly map out large-scale
structure.  Even with 8-10m-class telescopes, redshift surveys capable
of creating maps with resolutions of $\sim$ several Mpc out to $z \sim
1$ have been limited to elongated `pencil-beam' geometries over $\sim
1\,\sqdeg$ areas \citep[e.g.,][]{davis:2003, lilly:2007,
  le-fevre:2013}.  Similar maps out to $z \sim 2$ would require
volume-limited samples of $\mathcal{R} > 25$ galaxy redshifts
requiring very long ($\texp \sim 10\,\hrs$) exposures on 8-10m
telescopes in order to obtain secure redshifts from intrinsic
absorption lines\footnote{Galaxy redshifts up to $\mathcal{R} \leq
  25.5$ have been obtained with $\texp \sim 2-4\,\hrs$
  \citep[e.g.,][]{steidel:2004,lilly:2007}, but at the faint-end most of the successful redshifts were
  from objects that had emission lines. Obtaining redshifts from \emph{complete}
  samples that include non-emission line galaxies, require much longer
  exposures.}.  
Such surveys over cosmologically interesting volumes
would be extremely expensive on existing instrumentation, and will
likely be only feasible with future 30m-class telescopes. 

At $z \gtrsim 2$, an alternative probe of large-scale structure is the Lyman-$\alpha$ (\lya) forest absorption in the sightline of 
distant background sources \citep{lynds:1971}. 
With the insight that the \lya\ absorption in the photoionized intergalactic
medium (IGM) comes from residual \ion{H}{1} that directly traces the underlying dark matter 
overdensity \citep{cen:1994, bi:1995, zhang:1995, hernquist:1996, miralda-escude:1996, bi:1997}, 
the \lya\ forest has been established as the premier probe of the IGM and structure $z \gtrsim 2$ universe at density ranges 
near the cosmic mean. 
While the \lya\ forest has traditionally been treated as
1-dimensional probes of the IGM \citep{croft:1998, croft:2002, mcdonald:2000, mcdonald:2005, mcdonald:2006,
zaldarriaga:2003, viel:2004}, the BOSS \lya\ Forest Survey \citep{lee:2013} has recently measured \lya\ forest
correlations across multiple quasar lines-of-sight to constrain the large-scale baryon acoustic
oscillations (BAO) scale at $\langle z \rangle \approx 2.3$ \citep{busca:2013, slosar:2013, kirkby:2013}. 

The projected sightline density of the BOSS \lya\ forest quasars is $ \approx 17\, \persqdeg$ at a limiting
apparent magnitude of $\glim \approx 21.5$ \citep{bovy:2011,ross:2012,paris:2012}. 
While the number of quasars per square degree 
increases with limiting magnitude, the slope of the quasar luminosity function
is relatively flat \citep[e.g.,][]{palanque-delabrouille:2013} --- this is illustrated in Figure~\ref{fig:lumfunc}. 
The corresponding area density of quasars at limiting magnitudes of $\glim \approx 23$
is $\sim 50\,\persqdeg$ at $2.2 \lesssim z \lesssim 3$.
At around $g \sim 23$, however, star-forming Lyman-break galaxies 
(LBGs\footnote{Technically, `LBGs' refers to star-forming galaxies selected at $z\gtrsim 3$ by their 
$\lambda < 912\,\ang$ Lyman-limit absorption, but we use
the term to cover all $z\gtrsim 2$ galaxies with a star-formation far-UV continuum.}) begin to dominate the overall 
UV luminosity function (Figure~\ref{fig:lumfunc}). Furthermore, the LBG luminosity function \citep{reddy:2008} is considerably steeper than 
its quasar equivalent
--- at the $\glim \sim 24.5$ magnitude limit corresponding to $L_*$ galaxies at $z \sim 2.5$, 
the projected area density of LBGs is $\gtrsim 1000\,\persqdeg$ over $2.2 \gtrsim z \gtrsim 3$. 

\bfig
\epsscale{1.2}
\plotone{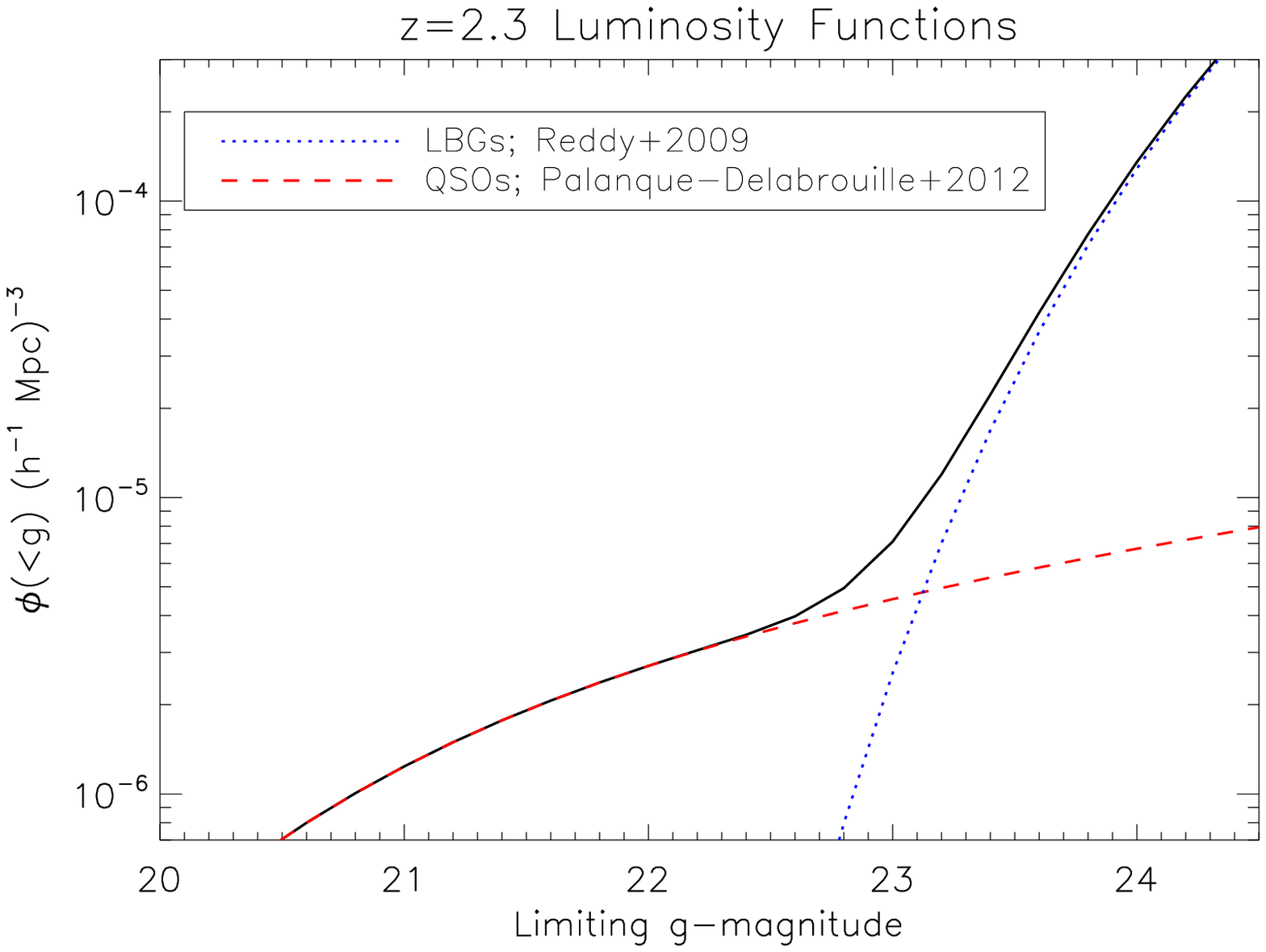}
\caption{\label{fig:lumfunc}
Luminosity functions used in this paper, shown here as a function of limiting $g$-magnitude. 
The red dashed curve shows the QSO luminosity function from 
\citet{palanque-delabrouille:2013}, while the dotted blue curve is the LBG luminosity function from 
\citet{reddy:2008}. The solid curve shows the sum of both luminosity functions. 
}
\efig

At sightline densities of $\sim 1000\,\persqdeg$, the typical projected separation between the sightlines is
$\sim 2\arcmin$; at $z \sim 2$, this corresponds to transverse comoving distances of $\sim 2\,\mpc$.
It then becomes possible to combine the transverse sampling with the line-of-sight absorption to generate
a 3D tomographic map of the \lya\ absorption field with a spatial resolution corresponding roughly to the sightline
separation\footnote{Although this can also be done with smaller sightline densities, e.g. 
M.\ Ozbek et al (in prep) is currently working on a tomographic map using the BOSS data.}.
Since each individual spectrum typically samples the \lya\ forest over $400-500\,\mpc$ along the line-of-sight, 
this method
has the potential to 
efficiently survey large volumes at $z \gtrsim 2$. 
\lya\ forest tomography is not a new idea: \citet{pichon:2001} presented a Wiener reconstruction formalism for
this problem, while \citet{caucci:2008} applied this formalism to numerical simulations to show that
the technique can recover the large-scale topology of the dark matter field at $z \sim 2$. 

It is often assumed 
that the spectra of $g \gtrsim 24$ LBGs required for \lya\ forest tomography cannot be be obtained with the 
current generation of telescopes, and will require 30m mirror apertures. 
In the literature, we have encountered rather formidable
observational requirements for IGM tomography, e.g. \citet{steidel:2009} specifies 
$\snr \sim 30$ per pixel at $R = 5000$ for $R = 24.5$ sources, while 
\citet{evans:2012} calls for $\mathrm{S/N} \ge 8$ per resolution element for $r=24.8$ sources. 
Such spectra would require exposure times of the order $\texp \sim 10$hrs even on 30m-class telescopes.
However, we could not find detailed justification for such steep requirements, neither for the resolution nor the
signal-to-noise.

However, according to the fluctuating Gunn-Peterson picture of the IGM \citep[e.g.,][]{croft:1997,rauch:1997,croft:1998}, 
\lya\ absorption is a non-linear tracer of the smoothly-varying dark matter distribution. Therefore, 
large-scale structure measurements on comoving scales of $\gtrsim 1\,\mpc$ do not have to resolve
individual \lya\ forest lines, and the individual spectra can be noisy if appropriate noise weighting
is implemented in the analysis \citep{mcdonald:2006,mcdonald:2007,mcquinn:2011}.
This is borne out by the success of the BOSS \lya\ Forest Survey, in which the typical 
 spectrum has $\snr \sim 2$ per angstrom within the \lya\ forest \citep{lee:2013}.

It is therefore the purpose of this paper to examine in detail the
actual requirements necessary to carry out \lya\ forest tomographic
mapping at spatial resolutions of $\sim 1-5 \,\mpc$.  We will first
evaluate the availability of absorption sightlines at various
magnitude limits and redshifts (\S~\ref{sec:nlos}).  In
\S~\ref{sec:tomosims}, we directly carry out tomographic
reconstructions on mock spectra derived from numerical simulations of the \lya\ forest, 
in which we have included instrumental effects reflecting various
choices of sightline density and exposure times.  As we shall show, 
\lya\ forest tomography on scales of $\sim 2-5\,\mpc$ is
already feasible on the current generation of 8-10m
telescopes with reasonable exposure times. We also discuss an analytic 
model for the reconstruction signal-to-noise (\S~\ref{sec:wiener_anal}), 
that allows us to quickly explore the 
effect of various exposure times and sightline densities on the reconstructed maps. 
In \S~\ref{sec:discussion} we will discuss possible science applications for 
\lya\ tomography and potential survey strategies on
various existing and proposed spectrographs.

We assume a flat $\Lambda$CDM cosmology with $\Omega_m=0.274$,
$\Omega_{\Lambda}= 0.726$ and $H_0 = 70 \,\kms \mathrm{Mpc}^{-1}$.

\section{Source Availability}
In this section, we first examine the availability of background sightlines for \lya\ forest tomography using
the published luminosity functions for LBGs and QSOs, followed by an analytic estimate of the observational
requirements for \lya\ forest tomography.

\subsection{Estimating Sightline Densities} \label{sec:nlos}
The critical observational parameter for 3D \lya\ forest tomographic mapping is the area density of effective background sightlines, \nlos. 
In the conclusions of their paper, \citet{caucci:2008}
tabulated the projected area density of $z \gtrsim 2$ background 
quasars and LBGs at several observed magnitude limits. 
However, this quantity is \emph{not} identical to \nlos, since the finite \lya\ forest path length in each background sightline
(typically $1040\,\ang \lesssim \lambrest \lesssim 1180\,\ang$) means that only a subset of these sightlines will pierce 
the volume at a given foreground redshift.

The effective number of sightlines per unit area that actually probe any given absorption redshift $z$
and limiting magnitude $\mlim$ is given by the differential sightline density
\begin{eqnarray} \label{eq:nlos}
\nlos(z, \mlim) = \int^{z_2}_{z_{1}} d\zbg& & \int^{\mlim}_{\infty} dm \frac{dl_c}{d\zbg}\, \phi(\zbg, m)
\end{eqnarray}
where $\phi(\zbg, m)$ is the luminosity function of background sources at redshift $\zbg$ and apparent magnitude
$m$, $dl_c$ is the comoving line-element along the line-of-sight. The integration limits on $\zbg$ take into account the
finite lengths of \lya\ forest in each sightline that could intersect the absorption redshift $z$:  $1+z_2 = (\lambalpha/\lambda_\mathrm{rest,min})(1+z)$ and $1+z_1 = (\lambalpha/\lambda_\mathrm{rest,max})(1+z)$
where $\lambalpha = 1215.7\,\ang$ is the \lya\ absorption wavelength and $[\lambda_\mathrm{rest,min},\lambda_\mathrm{rest,max}] \approx [1040\;\ang, 1180\;\ang]$ is the range of useful \lya\ forest wavelengths in each sightline.
This $1040\,\ang \leq \lambrest \leq 1180\,\ang$ range is chosen to avoid the quasar proximity zone as well as to avoid
having to predict the shape of the Ly$\beta$ emission line \citep{mcdonald:2006,lee:2013};
these criteria might not apply for LBGs, but we use it as a conservative value. %

\begin{figure} 
\epsscale{1.2}
\plotone{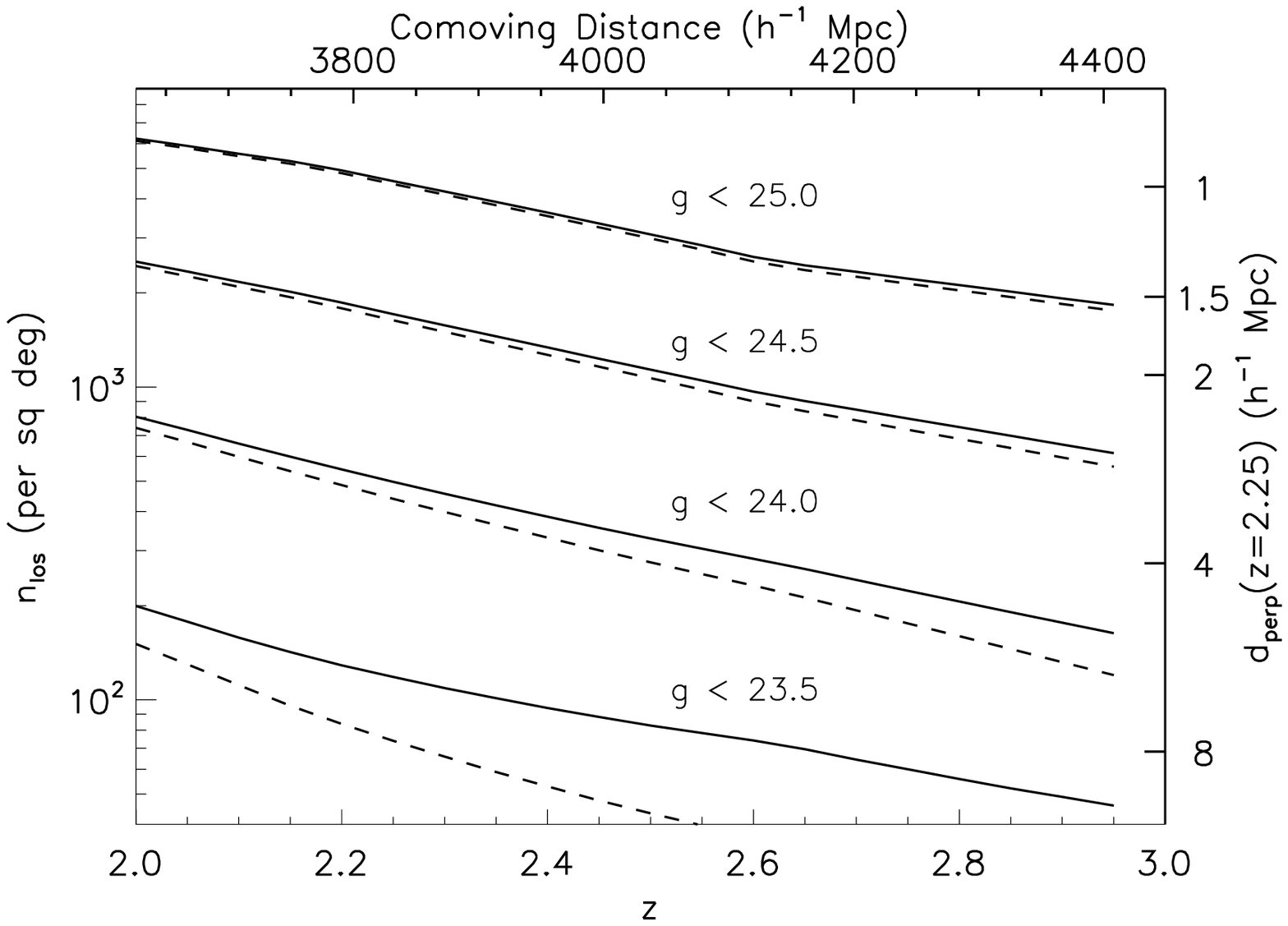}
\caption{\label{fig:nlos}
The number of background sightlines per sq deg, \nlos, available to probe redshift $z$ at several limiting
$g$-magnitudes. The dashed lines indicate the contribution from galaxies, while the solid line shows
the total contribution from both galaxies and quasars. The top axis indicates the comoving distance
to each redshift, while the right axis labels the typical sightline separation corresponding to \nlos, evaluated at
$z=2.25$. 
}

\end{figure}

We estimate \nlos\ by integrating over the published luminosity functions for both LBGs and QSOs,
illustrated in Figure~\ref{fig:lumfunc}. 
For the LBG contribution, we use the Schechter function fits from Table 14 in \citet{reddy:2008}, 
evaluated in the redshift range $1.9 < z < 2.7$.
This luminosity function was evaluated in
 $m_{\mathcal{RG}}$, a composite bandpass between $\mathcal{G}$ and $\mathcal{R}$ \citep[in the custom filter set of][]{steidel:1993} 
 that 
is roughly equivalent to Johnson $V$-band. 
In this paper we will work mostly in the SDSS $g$-band, so we make color corrections through the following:
we first generate a toy model of the mean LBG spectrum at $z=2.4$, represented by a power-law 
$f_\lambda \propto \lambda^{\beta}$ with \lya\ forest attenuation at $\lambda < 1216 (1+\zbg)$ given by \citet{becker:2013}, 
where $\beta = -1.1$ is the UV continuum slope estimated for $R \approx 24.5$ galaxies at $z=2.5$ from \citet{bouwens:2009}.
By comparing the convolution of this model spectrum by the various filter 
bandpasses and assuming $m_{\mathcal{RG}}=\sqrt{\mathcal{RG}}$, we find $g - m_{\mathcal{RG}} \approx 0.2$.
This appears consistent with the LBG color-color diagrams shown in \citet{steidel:2004}. 
The resulting \nlos\ distribution is shown as a function of $z$ by the dashed lines in Fig.~\ref{fig:nlos}.
At $z=2.45$, we find $\nlos = [48, 301, 1162, 3243]$ per sq deg at limiting magnitudes of $\glim = [23.5, 24.0, 24.5, 25.0]$, 
respectively, from galaxies. 

For the contribution of quasars to the background sightlines, we use
the quasar luminosity function published by \citet{palanque-delabrouille:2013} in
Equation~\ref{eq:nlos}.  The quasars provide $\nlos = [41, 54,71,91]$
per sq deg at $z = 2.45$ at $\glim = [23.5, 24.0, 24.5, 25.0]$.
Quasars clearly provide only a small fraction of background
sightlines at $\glim \ge 24$, but they represent the
brightest sources at $g \lesssim 23$ and cannot be neglected in an
actual survey.

The total \nlos\ from galaxies and quasars, at various limiting magnitudes, is shown by the solid curves in Fig~\ref{fig:nlos}. 
This illustrates the exponential increase in \nlos\ with limiting magnitude at
$g \gtrsim 23$: every magnitude increase in depth yields an order-of-magnitude more sightlines. 
Since \nlos\ quantifies the number of sightlines probing an infinitesimal redshift slice, 
one must observe a projected source density greater than \nlos\ to cover a finite distance along the line-of-sight at the same density, 
For example, to ensure roughly uniform $\nlos$ over the range $2.15 \leq z \leq 2.45$, a survey would have 
to target a projected source density of $\sim 1.8\, \nlos(z=2.45)$, with sources spanning $2.3 \lesssim \zbg \lesssim 3.1$. 
In this paper, we will adopt this scenario as our fiducial survey, i.e. a $\langle z \rangle = 2.25$ survey 
that assumes a constant differential sightline density of $\nlos(z=2.45)$ over the entire volume, but all other parameters (e.g.\ angular diameter distance)
are evaluated at the lower redshift. Since the comoving distance between $z=2.15$ and $z=2.45$ is $\approx 250\,\mpc$, 
such a survey would cover a volume of roughly $1.1\times10^{6}\,\mpccube$ per square degree of sky observed.

For a given \nlos, we can calculate the typical inter-sightline separation, $\dperp \approx \sqrt{1/\nlos}$.
This can be written in units of comoving transverse Mpc:
\beq \label{eq:dperp}
\dperp \approx \;\left[\frac{\nlos}{4200\,\persqdeg}\right]^{-1/2} \left[\frac{1+z}{3.25} \right]^{-3/2}\,\mpc,
\eeq
where the redshift dependence comes from changes in angular diameter distance.
Intuitively, we do not expect to be able to map out scales 
much smaller than \dperp, so
 \dperp\ na\"ively sets the spatial resolution \sigthreed\ of the tomographic reconstructions feasible from a given set of
 sightlines. However, as we shall see later, the scale factor relating \sigthreed\ and \dperp\ is of order unity, 
 but can be varied depending on the desired map properties.
The right-axis of Figure~\ref{fig:nlos} labels \dperp\ corresponding to several values of \nlos,
assuming the angular diameter distance at $z=2.25$. 
In terms of limiting magnitudes, a survey needs to reach $\glim = [23.5, 24.0, 24.5, 25.0]$
in order to achieve source densities such that $\dperp = [7.1, 3.6, 1.9, 1.2]\,\mpc$.

We find that a simple analytic approximation, 
\beq \label{eq:nlos_fit1} \log_{10}\nlos \approx  \glim - 21.5 - 0.75(z-2.45) \eeq 
fits the source counts
to within $\sim 15\%$ over the redshift and limiting magnitudes considered here.
This also allows us to write down the median $g$-magnitude corresponding to \nlos:
\beq \label{eq:nlos_fit}
\bar{g} \approx \log_{10}\nlos + 21.0 + 0.75(z-2.45),
\eeq
by simply setting $\nlos \rightarrow 0.5\,\nlos$ in Equation~\ref{eq:nlos_fit1}.

 Alternatively, we insert the approximation of Equation~\ref{eq:dperp} to find
\begin{eqnarray} \label{eq:dperp_fit}
\log_{10}\dperp &\approx& -\frac{1}{2} [\glim - 25.1 - 0.75(z-2.45)] \nonumber \\
 &\approx& -\frac{1}{2} [\bar{g} - 24.6 - 0.75(z-2.45)].
\end{eqnarray}
We will use these approximations in the simple estimates of the following sub-section.

\subsection{Order-of-Magnitude Exposure Time Estimates} \label{sec:texp_anal}
We will now make some simple analytic estimates of the observational requirements
required to carry out IGM tomography at various scales, \sigthreed. While we will carry out definitive tests
with simulations in the next section, these simple calculations argue that \lya\ tomography should already
be feasible on the current generation of 8-10m telescopes.

The first parameter we consider is the spectral resolving power, $R \equiv \Delta \lambda / \lambda$, 
required for IGM tomography. We conjecture that not only is it unnecessary to resolve individual 
\lya\ forest absorbers (requiring echelle spectrographs with $R\gtrsim 10^4$), it is necessary only
to resolve the desired 3D reconstruction scale of the map, which
we define as $\sigthreed$.
By converting \sigthreed\ to its equivalent 
span in observed wavelength\footnote{For reference, $1\,\mpc$ comoving distance spans $1.25\,\ang$ or $\Delta v = 95\,\kms$
along the line-of-sight at $z=2.25$}, we find
\begin{eqnarray} \label{eq:res}
R  &>& 1300 \;\left(\frac{1\,\mpc}{\sigthreed}\right)\;\left[\frac{(1+z)}{3.25}\right]^{-1/2}. 
\end{eqnarray}
This implies that only moderate-resolution spectra 
are required for IGM tomography, with up to
$R\approx 1300$ required for $\sigthreed=1\,\mpc$ reconstructions.
Alternatively, if one desires $\sigthreed \approx 3 \,\mpc$ maps then $R\approx 400$
is sufficient --- this is well within the resolving power of the main moderate-resolution blue grisms on VLT-VIMOS 
\citep[HR-Blue;][]{le-fevre:2003}
and Keck-LRIS \citep[B600/4000;][]{oke:1995}.

We now investigate the telescope exposure time, \texp, required to reach a given sightline density and transverse separation.
In the background-limited regime, the relationship between \texp\ and source magnitude, $g$, is given by
\beq \label{eq:texp}
\texp = [\snr]^2 10^{0.8 (g - g_0)}\; \mathrm{hrs}, 
\eeq
where $\snr$ is the spectral signal-to-noise per angstrom, while $g_0$ is the magnitude at which a telescope and instrument
combination gives $\snr = 1$ per angstrom 
with an integration time of $\texp  = 1\,\mathrm{hr}$. 
We estimate $g_0 = 24.75$ for the VIMOS spectrograph on the VLT using the 
ESO exposure time calculator\footnote{\url{http://www.eso.org/observing/etc/}}.
This assumed $0\arcsec.8$ seeing, $1\arcsec$ slit size, 1.3 airmass
 and 3 days from new moon on the $R=1150$ HR-Blue grating. 
 We also assume a point source, since the typical $z\sim 2-3$ LBGs are compact objects with 
 half-light radii of $\sim 0\arcsec.3$ \citep[e.g.,][]{shapley:2011}; 
furthermore, \citet{giavalisco:1996} found that for 70\% of the LBGs in their sample, 
 the magnitudes within an $0\arcsec.7$ aperture differed from the isophotal magnitude by less than $0.1$ mag.

Using Equation~\ref{eq:dperp_fit}, we can relate the median source magnitude to the expected separation between sightlines, and find
\begin{eqnarray} \label{eq:texp_dperp}
\texp &=& 49 \left( \frac{\snr}{8 \:\mathrm{per\: \ang}}\right)^2 \left( \frac{1\,\mpc}{\dperp} \right)^{1.6} \nonumber \\
	& & \hphantom{yoloyoloyoloyo} \times\: 10^{0.8(24.75-g_0)}\, \mathrm{hrs}.
\end{eqnarray}
This indicates that $\texp = 49\,\mathrm{hrs}$ would be required to obtain 
absorption spectra with $\snr = 8$ per angstrom \citep[roughly the value quoted by][as necessary for IGM tomography]{evans:2012} 
from the $\bar{g} \approx 24.6 $ sources that are separated by $\dperp \approx 1\,\mpc$
between lines-of-sight. 

If we assume that the spatial resolution of a tomographic map, $\sigthreed$, is given by the 
sightline separation $\sigthreed \approx \dperp$, then $\sigthreed = 1\,\mpc$ tomography
is clearly not feasible with current 8-10m telescopes, and would require the $\sim 15-20\times$ greater collecting 
areas of 30m-class telescope to achieve the necessary depths.
However, due to the $\texp \propto \dperp^{-1.6}$ scaling, the requisite exposure times drop quickly for
 maps with coarser resolution: to reach $\dperp = 4\,\mpc$ at fixed signal-to-noise, we find $\texp = 5.3\,\hrs$. Exposure 
times of this order are regularly carried out in galaxy redshift surveys on 8m telescopes, e.g.\ 
the zCOSMOS-Deep survey \citep{lilly:2007}. However, the data from these surveys are not useful for \lya\ tomography
since they are usually low-resolution ($R \sim 200$) spectra
that do not have sufficient resolution to resolve structure along the line-of-sight (Equation~\ref{eq:res}). 
 
However, it is still unclear what spectral signal-to-noise ratio is required for \lya\ tomography ---
in the previous paragraph we have picked a somewhat large value assumed by \citet{evans:2012}. 
This is an important question to address, as the exposure time depends sensitively on the minimum signal-to-noise requirement 
of the survey. For example, if $\snr = 4$ per angstrom were the requirement, then only $\texp\approx 12 \,\hrs$
would be required to reach a source separation of $\dperp = 1\,\mpc$.
Moreover, we have assumed
that the spatial resolution of the map is directly set by the typical sightline separation, $\sigthreed = \dperp$.
This relationship makes sense at a rough intuitive level since we do not expect to resolve features on scales
smaller than the sightline separation, but we need to investigate the exact relationship between $\sigthreed$ and
\dperp, which is possibly also scale-dependent. 
To address these questions, we will, in the next section, directly carry out tomographic reconstructions on mock \lya\ forest
spectra derived from numerical simulations.

\section{Tomographic Reconstruction on Simulations} \label{sec:tomosims}
In this section, we use mock skewers derived from numerical simulations to explicitly test the signal-to-noise
and sightline densities required to map out the IGM at various scales.
We first describe the simulations and reconstruction technique, and then discuss the quality of the reconstructions as a function
of telescope exposure time and reconstruction scale. Finally, we will derive an analytic expression that allows us to understand
some of the trends we see in the simulated reconstructions.

\subsection{Simulations and Mock Spectra} \label{sec:sims}

We use the dark matter-only numerical simulation used in
\citet{white:2010b}. Briefly, this simulation was run using 
 a TreePM code \citep{white:2002}
that evolved $2048^3$ dark matter particles over a $(250\,\mpc)^3$
periodic cube. The resulting particle masses are $1.4\times10^8
\,h^{-1} M_\odot$ and have a Plummer equivalent smoothing of
$2.5\,\kpc$. The assumed cosmological parameters were $\Omega_m =
0.274$, $\Omega_\Lambda = 0.726$, $h=0.7$, and $\sigma_8 = 0.8$.  Our
particular epoch of interest is $z\approx 2.3$, so we used the
simulated dark matter distribution output at $z=2.57$ but subsequently
treat it as if it were at $z=2.25$, so that values for e.g.\ the mean
forest transmission, sightline densities, and angular diameter
distance are computed for the latter redshift --- for the purposes of
this paper the deviations from the true simulation redshift are irrelevant.
From the simulated particle distribution, we generate simulated \lya\ forest 
spectra following the procedure described
in \citet{rorai:2013}; but see also \citet{le-goff:2011, greig:2011}, and \citet{peirani:2014} for alternative methods of generating
simulated \lya\ forest skewers in the context of large-scale structure studies. 

In brief, the dark matter field is first smoothed by an assumed Jeans scale, $\lambda_J$, to mimic
the Jeans pressure smoothing experienced 
by baryons, followed by the use of the
fluctuating Gunn-Peterson approximation \citep[e.g.,][]{croft:1998,gnedin:1998,weinberg:2003} to link the underlying dark matter 
overdensity $\Delta_\mathrm{dm} \equiv \rho_\mathrm{dm}/{\langle {\rho}_\mathrm{dm}} \rangle$
to the \lya\ optical depth, $\tau \propto \Delta_\mathrm{dm}^{2-0.7(\gamma-1)}$, where $\gamma$ governs the (assumed) power-law
temperature-density relationship of the IGM, $T(\Delta) \propto \Delta^{\gamma-1}$. 
The Jeans scale is not well-constrained by observations but we choose $\lambda = 100\,\kpc$ which conforms
to theoretical expectations \citep{gnedin:1998}.
Meanwhile, the value of the temperature-density slope has been the subject of multiple discrepant measurements
\citep{bolton:2008, viel:2009, garzilli:2012, calura:2012,rudie:2012, lee:2014} so we pick $\gamma=1.0$ as a 
meta-average from these papers. 
The uncertainties in these (and other) IGM parameters are unimportant for us in this paper since we will be
making quantitative comparisons only between the `true' and tomographically reconstructed absorption fields, 
but future efforts to directly the infer the dark-matter distribution will have to marginalize over uncertainties in the 
astrophysics of the IGM, which would otherwise lead to errors in the contrast of the resulting maps.

The optical depths are next convolved with the peculiar velocity field to give the skewers in redshift space.
The \lya\ forest transmission is then given by $F = \exp(-\tau)$; the full set of skewers is normalized
to give a mean transmission, $\langle F \rangle$, consistent with the measurements of
 \citet{faucher-giguere:2008}.
In total, there were 62500 skewers covering the full $250^3 \,\mpccube$ simulation box 
at transverse separations of $1\,\mpc$, with each skewer binned into 2048 line-of-sight pixels with a size of $12\,\kms$.

For a given tomographic reconstruction, we randomly select a number of sightlines from our full
$1\,\mpc$ grid until we reach the desired area density \nlos. 
Note that due to the quantized nature of our sightlines in the transverse plane, we limit ourselves 
to reconstructions where the inter-sightline separation is $\dperp > 1.4\,\mpc$.
We then smooth the spectra to the assumed spectrograph resolution 
using a 1-dimensional Gaussian kernel.
In practice, we find no difference in the reconstruction quality at different resolutions, so long as
map resolution, $\sigthreed$, is resolved by the spectrograph 
 (i.e.\ Equation~\ref{eq:res} is obeyed).

Assuming a complete target selection, the luminosity functions allows us to randomly assign 
an object magnitude $g$ to each sightline, giving us a distribution of source magnitudes down to 
the survey limit $\glim$ that corresponds to the chosen sightline density \nlos.
The telescope exposure time, \texp, then relates the assigned magnitude, $g$,
of each source to the 
 signal-to-noise in the \lya\ forest via Equation~\ref{eq:texp}. 
The quantity $g_0$ in that equation is determined by the specific telescope and
spectrograph; we use as a fiducial value $g_0 = 24.75$ corresponding to the HR-Blue mode of the VIMOS
spectrograph on the VLT, although it is easy to rescale \texp\ to a different telescope collecting area and
instrument throughput.

\begin{figure}
\epsscale{1.2}
\plotone{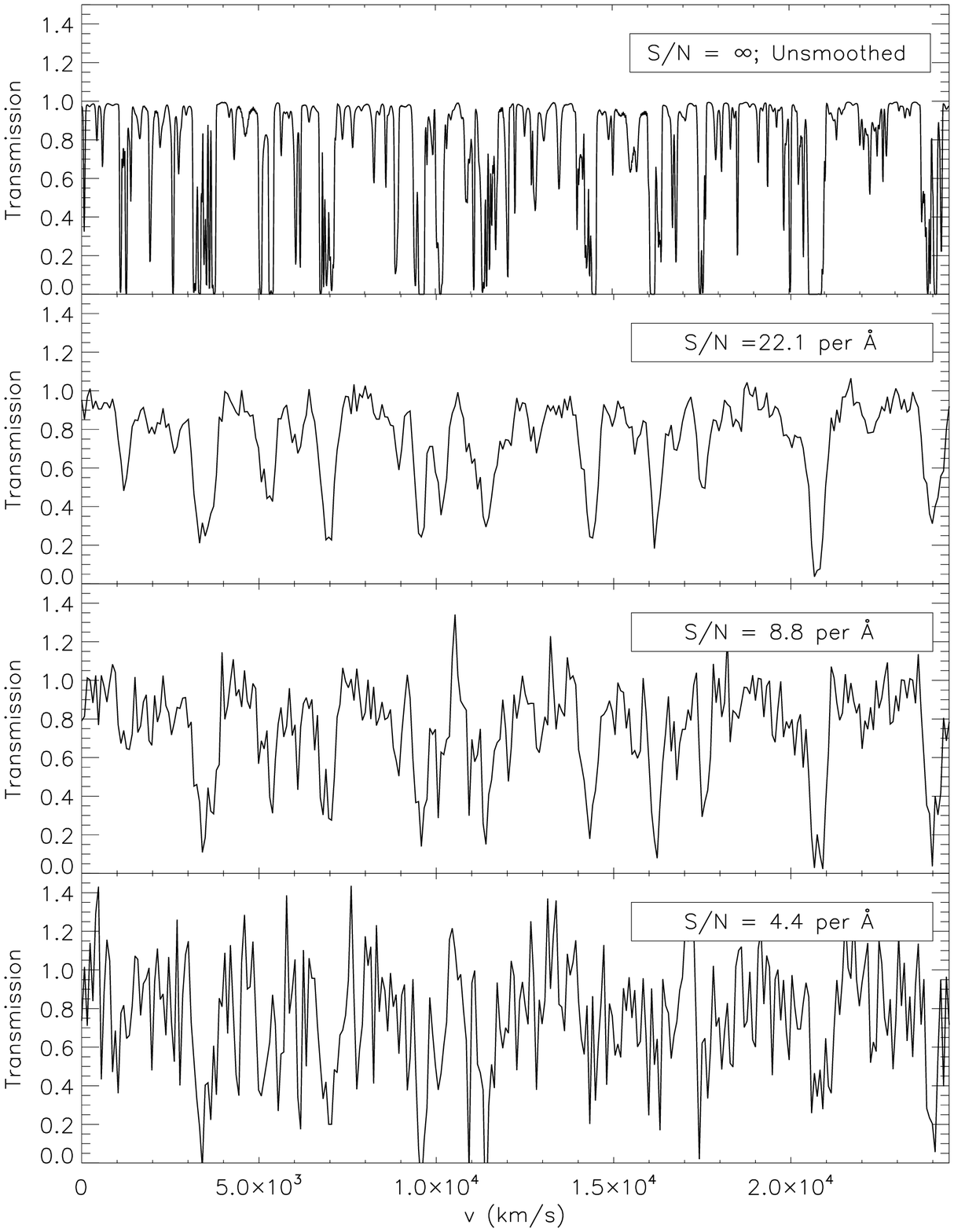}
\caption{\label{fig:noisy_spectra}
Mock spectrum from our simulations, shown at several signal-to-noise ratios. 
The top panel shows the spectrum without noise or resolution effects, and binned in
$11\,\kms$ pixels. 
The lower 3 panels show the same spectrum after it has been smoothed to mimic a resolving power of 
$ R=1150$, rebinned to $1\,\ang$ pixels, and had noise added corresponding to the signal-to-noise
shown in the labels. The lower panel corresponds roughly to the signal-to-noise limit set by the necessity
to make a redshift identification. Assuming $\texp = 5\,\hrs$ on the VIMOS, the noisy spectra correspond to
source magnitudes of $g = [22.3, 23.3, 24.0]$ from top to bottom.
}
\end{figure}

For each mock spectrum we then add mock noise to the native simulation pixels by adding a vector of
Gaussian deviates with a standard deviation equal to $(\sqrt{95/12})/\snr $,
where $\snr$ is determined by \texp\ and individual source
magnitude, and the factor in square-roots are to rescale the noise from the $95\,\kms$ velocity
separation corresponding to $1\ang$ to the $12\,\kms$ simulation pixels.  
Figure~\ref{fig:noisy_spectra} shows several examples of
our mock spectra that have been smoothed to the assumed spectrograph resolution, 
rebinned to $1\,\ang$ pixels, and have had pixel noise
added, assuming source magnitudes of $g = [22.3, 23.3, 24.0]$ and an exposure time of $\texp=5$ hrs on the VIMOS.

\subsubsection{Wiener Reconstruction} \label{sec:wiener}

To carry out the tomographic reconstructions, we have written a parallel Fortran90 code that closely follows
 the Wiener interpolation algorithm described in \citet{caucci:2008}. This code, which will be described in more detail
 in M.\ Ozbek et al (in prep), was written to create large-scale ($\sigthreed \sim 20\,\mpc$) tomographic
 maps from the BOSS \lya\ forest data.
In this approach, the reconstructed map $m$ is estimated from the data $d$ by evaluating
\begin{eqnarray} \label{eq:wiener}
m &=&  \mathbf{K} \cdot d \nonumber \\
                  &=&         \cmd\cdot (\cdd + \mathbf{N})^{-1} \cdot d, 
\end{eqnarray}
where $\mathbf{K} \equiv \cmd\cdot (\cdd + \mathbf{N})^{-1}$ is a form of Wiener filter \citep{wiener:1942,press:1992,zaroubi:1995}
and $\mathbf{N}$ is the noise covariance matrix (assumed to be diagonal), 
while $\cmd$
and $(\cdd + \mathbf{N})$ describe the map-data and data-data covariances, respectively. 
In principle, $\cmd$ and $\cdd$ are set by the observed flux power spectrum or correlation function, 
but it is well-known that for the purposes of constructing the Wiener filter these only need to be approximately correct \citep[see, e.g,][]{press:1992}.
In this paper we therefore adopt the \emph{ad hoc} approach of \citet{caucci:2008}, which assumes 
$\cdd = \cmd = \mathbf{C(r_1,r_2)}$ and
\begin{eqnarray} \label{eq:corrfunc}
 \mathbf{C(r_1,r_2)} &=& \sigma_F^2  \exp\left[-\frac{(\Delta r_\parallel)^2}{L^2_\parallel}\right] \exp\left[-\frac{(\Delta r_\perp)^2}{L^2_\perp}\right],
\end{eqnarray}
where $\Delta r_\parallel$ and $\Delta r_\perp$ are the distance between 
$\vec{r_1}$ and $\vec{r_2}$ along, and transverse, to the line-of-sight, respectively. 
In the case of $\cmd$, $\vec{\Delta r}$ corresponds to the separation between 
a grid point in the final map grid and a pixel in one of the 
sightline skewers, while for $\cdd$ the $\vec{\Delta r}$ is between two separate pixels in the absorption skewers, 
whether in different skewers or along the same skewer. 
The variance of the flux fluctuations, $\sigma_F^2$ is, for the reconstructions, measured directly from the 3D \lya\ forest field 
of the simulation itself. 
The parameters $L_\parallel$ and $L_\perp$ set the correlation length in the line-of-sight and transverse directions. 
We set $L_\parallel$ to the line-of-sight distance corresponding to the smoothing FWHM of the assumed instrumental
resolution (Equation~\ref{eq:res}), 
while $L_\perp = \dperp=\sqrt{1/\nlos}$ as suggested by \citet{caucci:2008}.

In practice, we first convert the \lya\ flux transmission $F=\exp(-\tau)$ skewers to fluctuations about the mean-flux,
$\delta_F = F /\langle F \rangle -1 $. 
The $\delta_F$ from the smoothed and noisy mock spectra are then binned into 
a $50^3$ grid. 
Since the algorithm is computationally expensive, we carry out reconstructions on subvolumes of 
$150^3\,\mpccube$, $100^3\,\mpccube$, or $50^3\,\mpccube$ extracted from the full simulation box and binned to $50^3$ grid cells.
The exact choice of subvolume depends on the tomographic reconstruction scale \sigthreed,
but we ensure that the grid cell sizes are always $\leq \sigthreed$. We have checked that the choice of
binning does not significantly affect our subsequent results.
The binned fluctuations $\delta_F$ constitute the `data' matrix $d$ in Equation~\ref{eq:wiener}.
Similarly, the noise matrix $\mathbf{N}$ is estimated for each cell by summing in quadrature the
inverse of the noise variances, $\sigma^2_N$, of the contributing sightlines.  
This step ensures correct weighting based on the signal-to-noise, and is crucial for a 
reasonable reconstruction since our sightlines are always dominated by the faint-end of the luminosity function.

To speed up the reconstruction, the matrix inversion in Equation~\ref{eq:wiener} is carried out separately
on chunks of $5^3$ gridcells. Buffer regions of $\sim 2\sigthreed$ are added to each face of the chunks to 
mitigate edge artifacts; we discard these buffer zones from the final reconstructed volume. 
As a final step, the reconstructed volume is smoothed by a 3D Gaussian kernel with standard deviation
$\sigthreed$. \citet{caucci:2008} set $\sigthreed = 1.4\dperp$, but we shall treat it as a free parameter that
allows us to vary the mapping scale and reconstruction fidelity of the map, as we shall 
see in the next section.

\subsection{Results}

 \begin{figure*}
\centering



\vspace{12pt}
\begin{overpic}[width=\textwidth,unit=1pt]{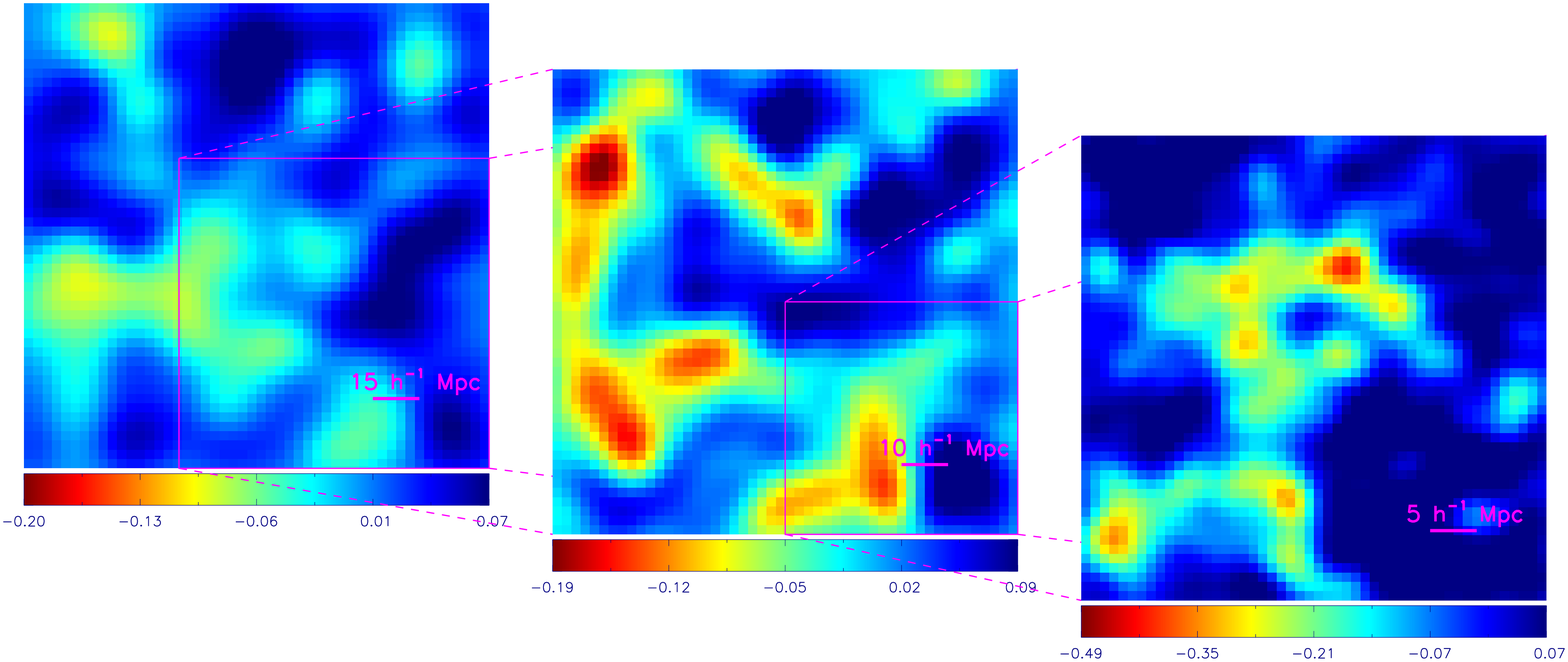}
\put(10, 220){\textsf{\small $\mathsf{n_{los} = 225\, deg^{-2}}$, $\mathsf{\epsilon_{3D} =7.1\,h^{-1} Mpc}$}}
\put(183, 200){\textsf{\small $\mathsf{n_{los} = 504\, deg^{-2}}$, $\mathsf{\epsilon_{3D} =4.0\,h^{-1} Mpc}$}}
\put(356, 179){\textsf{\small $\mathsf{n_{los} = 2218\, deg^{-2}}$, $\mathsf{\epsilon_{3D} =1.4\,h^{-1} Mpc}$}}
\end{overpic} \\
\vspace{-17pt}
\includegraphics[width=\textwidth]{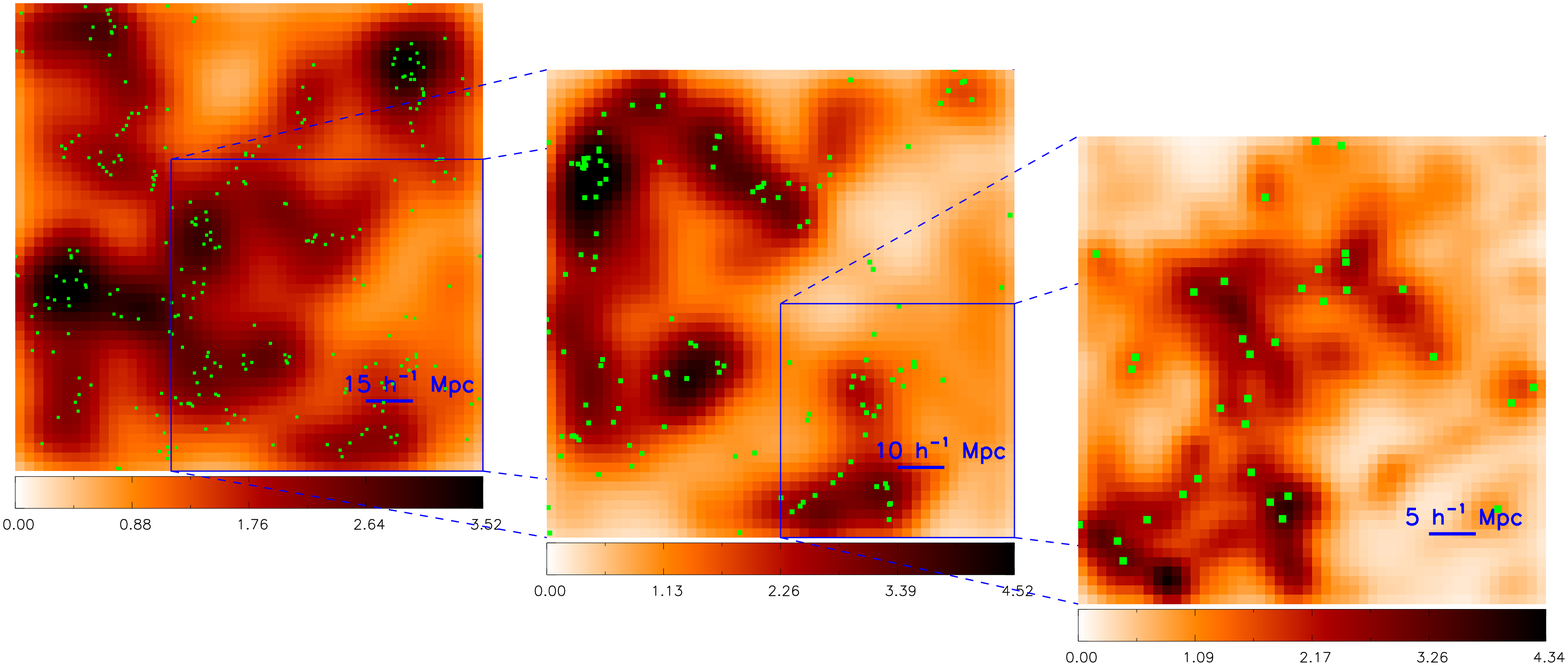}

	\caption{\label{fig:recon}
	Simulated \lya\ forest tomographic reconstructions at various scales $\sigthreed$, shown in the plane of the sky. 
	The top-row shows the \lya\ absorption field reconstructed
	from randomly-selected sets of mock spectra with various areal densities $\nlos$ and smoothed to $\sigthreed$ (both labeled on top),
	while the bottom row shows the corresponding dark matter overdensity smoothed to the same $\sigthreed$. 
	From left to right, the dimensions of the simulation slices are 
	$(150\,\mpc)^2 $, $(100\,\mpc)^2$, and $(50\,\mpc)^2$ respectively with thickness of $3\,\mpc$ along the line-of-sight; 
	the smaller slices are subvolumes of the larger slices. 
	These maps correspond to transverse areas of $[5.09\,\mathrm{deg}^{-2}, 2.26\,\mathrm{deg}^{-2}, 
	0.56\,\mathrm{deg}^{-2}]$ on the sky (from left to right), 
	and we have assumed
	telescope exposure times of $\texp = [3, 5, 16]\,\hrs$ to generate the noisy mock spectra. 
	The green dots overlaid on the DM maps show a complete sample of 
	$\mathcal{R} \leq 25.5$ galaxies co-eval with the \lya\ forest field, obtained from abundance matching to dark-matter halos in the simulation
	volume. 
	}
\end{figure*} 

In this section, we carry out tomographic reconstructions on the simulated \lya\ forest absorption skewers described
in \S~\ref{sec:sims}, in order to study the quality of the resulting maps, at various spatial resolutions \sigthreed, as a function of 
observational parameters.

The primary observational parameters we can vary are the differential sightline density, $\nlos(z)$, 
that is tied to the limiting magnitude\footnote{We ignore, for now, inefficiencies in target selection and 
assume that we can observe all
background sources down to the magnitude limit.}, $\glim$, 
and ultimately the telescope exposure time, \texp. 
In principle, \nlos\ and \texp\ can be varied independently but
we begin by assuming that \texp\ is tied to $\glim$ and $\nlos$. 
In other words, longer exposure times yield more sightlines because more sources become available at
fainter magnitudes. While it is in principle possible to target increasingly noisy faint sources below a given magnitude limit,
  there is, in practice, a minimum spectral signal-to-noise threshold below which a redshift identification becomes difficult.
For LBGs with no intrinsic \lya\ emission line, $\snr=4$ per angstrom seems to be a 
conservative threshold for measuring the redshift of a 
non-emission line LBG from its intrinsic absorption lines \citep{steidel:2010}.
Combining Equations~\ref{eq:nlos_fit1} and \ref{eq:texp}, we therefore have
\begin{eqnarray} \label{eq:texp_nlos}
\texp &\approx& 9  \left(\frac{\nlos}{1000\,\persqdeg}\right)^{0.8} \,\mathrm{hrs} \nonumber \\
&\approx& 9 \left(\frac{2.1\,\mpc}{\dperp} \right)^{1.6} \,\mathrm{hrs}
\end{eqnarray}
where we have assumed the performance of the HR-Blue mode of the VIMOS spectrograph on the VLT.
This equation is only an approximation for illustrative purposes, because for the simulated spectra we directly use the source luminosity functions 
(\S~\ref{sec:nlos}) to determine the relationship between $\glim$ and $\nlos$. 

For a given \nlos, we randomly select simulated sightlines and assign to each sightline a source magnitude
(assuming the luminosity functions described in \S~\ref{sec:nlos}) and seed the pixels with Gaussian random
noise assuming a signal-to-noise given by $\texp$ (Equation~\ref{eq:texp}).
We then carry out the Wiener filtering procedure described in the previous section. 
The final step involves smoothing the reconstructed volume with a 3D Gaussian with a 
standard deviation
\sigthreed\ --- we treat this as a free parameter that allows us to choose the final map resolution.

In Figure~\ref{fig:recon} we present transverse-plane slices of simulated \lya\ forest tomographic reconstructions
with smoothing scales of $\sigthreed=[7.1, 4.0, 1.4]\,\mpc$, generated from mock surveys with
$\nlos = [225, 504, 2218]\,\persqdeg$ and $\texp=[3, 5, 16]\,\hrs$ (we will explain the choices of \sigthreed\ later).
The reconstructions were carried out on sub-volumes of the overall simulation box, with dimensions  
$(150\,\mpc)^3 $, $(100\,\mpc)^3 $, and $(50\,\mpc)^3$ 
for the $\sigthreed=[7.1, 4.0, 1.4]\,\mpc$ reconstructions, respectively; 
the smaller boxes are sub-volumes of the larger boxes.
This is reflected in the slices shown in  Figure~\ref{fig:recon}; 
for example, one can see how the overall anvil-like structure in the $\sigthreed=1.4\,\mpc$
map (at right) varies as we move to coarser map resolutions. 

For a qualitative comparison, 
we juxtapose the underlying dark matter overdensity field 
$\Delta_\mathrm{dm} \equiv \rho_\mathrm{dm}/\langle \rho_\mathrm{dm} \rangle$, 
clipped to $\Delta_\dm \leq 20$ and smoothed to the same \sigthreed\ as the corresponding IGM maps.
The tomographic reconstructions agree very well with the underlying DM overdensity: 
features on scales larger than several $\sigthreed$ 
are well recovered, although the contrast 
is somewhat different due to the different dynamic range of the exponential $F\equiv \exp(-\tau_\mathrm{Ly\alpha})$
transformation in the \lya\ forest.
It is also clear that, even at the coarsest reconstruction scales, IGM tomography is exceptionally
good at detecting voids in large-scale structure, a non-trivial endeavour with galaxy redshift surveys even at low-redshifts
\citep[e.g.,][]{tinker:2008,kreckel:2011,sutter:2012}.
The efficient detection of voids will facilitate the study of the topology of large-scale structure out to high-redshifts.
However, in this paper we do not move beyond this qualitative comparison between the tomographic absorption maps 
and the dark-matter field; we will address the problem of inverting the dark-matter field to a future paper.

It is important to emphasize that these maps were reconstructed from mock spectra that \emph{include realistic 
pixel noise assuming feasible exposure times on existing 8-10m telescopes}. The 5-hour exposure times, 
which enables $\sigthreed=4\,\mpc$ maps (middle column of Figure~\ref{fig:recon}), is already regularly carried out for galaxy redshift surveys on 8m telescopes --- indeed,
we have set \texp\ by the necessity of measuring the source redshifts. 
The longer $\texp \gtrsim10\,\hrs$ exposure times required for $\sigthreed \lesssim 2\,\mpc$ tomography 
is at the margins of integration times that have been attempted for individual pointings in a galaxy redshift survey.
However, the stunning detail revealed by these maps is arguably unparalleled by any other cosmographical technique
outside of the $z \lesssim 0.1$ Local Universe, 
and should motivate attempts to carry this out on existing instrumentation.

\bfig
\plotone{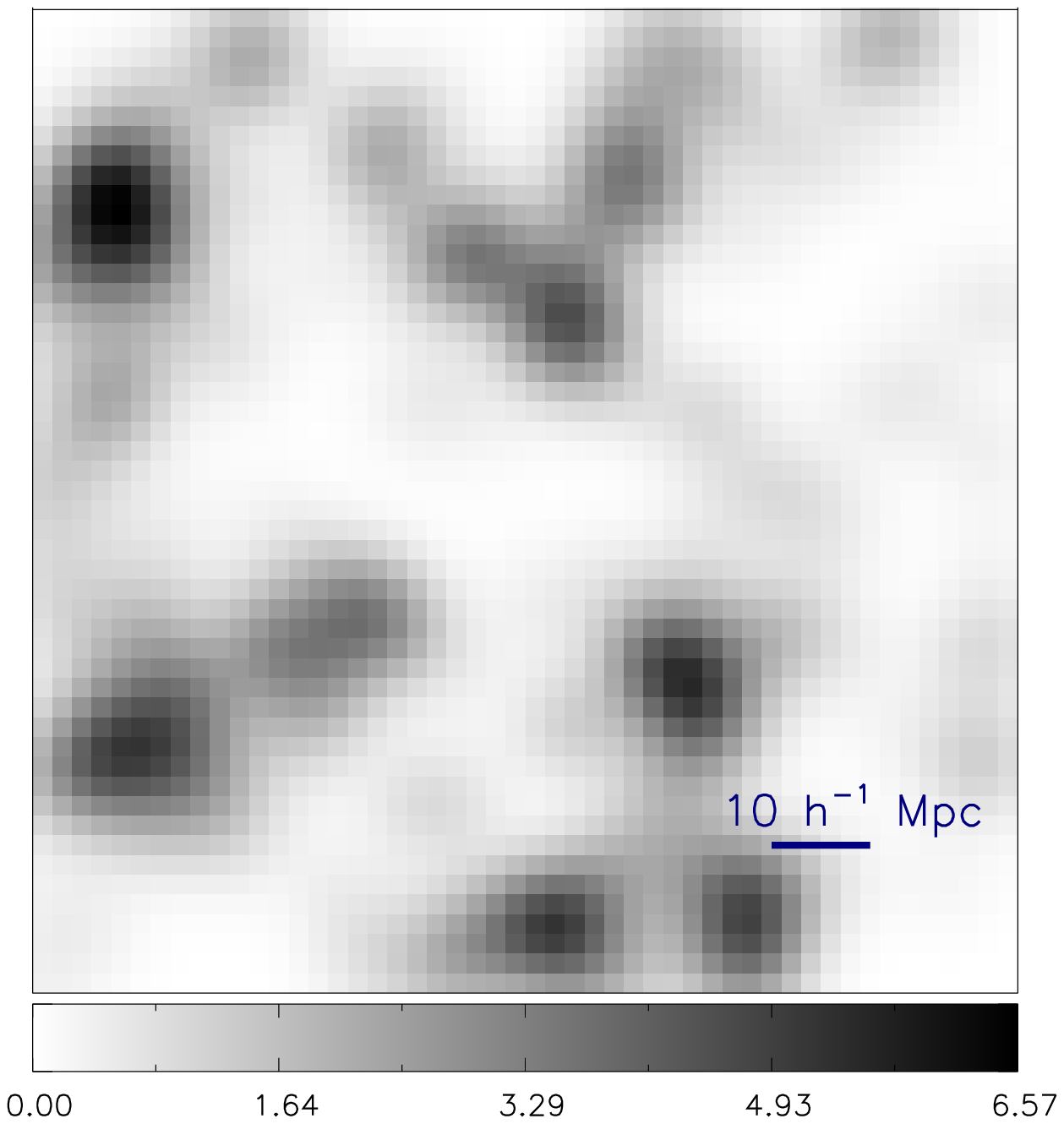}
\caption{\label{fig:galmap} 
Galaxy overdensity field in the same simulated slice shown in the middle panels of Fig.~\ref{fig:recon}. 
This used the distribution of $\mathcal{R} \leq 25.5$ galaxies obtained through halo abundance matching with the dark-matter
halos in the simulation, which was then smoothed
with a Gaussian filter with $4\,\mpc$ standard deviation to match the $\sigthreed$ of the tomographic reconstructions
in Fig.~\ref{fig:recon}. The dimensions of this map is $100\,\mpc \times 100\,\mpc$ 
in the plane of the page, and $2\,\mpc$ into the plane of the page. 
}
\efig

To compare with the tomographic maps, we have also generated a mock galaxy catalog through halo abundance-matching
\citep{kravtsov:2004,tasitsiomi:2004,vale:2004}
between the DM halo catalog from the simulation and the \citet{reddy:2008} LBG luminosity function. On the
DM overdensity maps we have over-plotted the positions of $\mathcal{R} \leq 25.5$ galaxies that fall within the same
simulated volume. 
These are foreground galaxies that would not be targeted as background sources within a tomographic survey, 
but sub-samples of such galaxies could be obtained through other surveys for comparison with the tomographic maps 
(or vice-versa: the tomographic mapping survey could target well-studied galaxy fields such as CANDELS, 
\citealt{grogin:2011,koekemoer:2011}, or 3D-HST, \citealt{brammer:2012}).
It is clear that the LBGs are clustered around the same features traced out by the 
 dark matter distribution and the \lya\ forest absorption, suggesting that \lya\ tomography will be a powerful tool in the study of galaxy environments at $z \sim 2$.

In Figure~\ref{fig:galmap}, we plot the overdensity of $\mathcal{R} \leq 25.5$ galaxies, 
$\Delta_\mathrm{galaxy}= \rho_\mathrm{galaxy}/\langle \rho_\mathrm{galaxy} \rangle  $ occupying the same simulation slice 
as the $\sigthreed=4\,\mpc$ tomographic reconstruction shown in the middle panel of Figure~\ref{fig:recon}.
This has been smoothed with a $\sigma=4\,\mpc$ Gaussian kernel to match the \lya\ forest reconstructions.
The strongest galaxy overdensities form quasi-spherical groups that trace out highly overdense regions of the 
simulation volume, but do not trace well the filamentary structures closer to mean-density.
This is further illustrated by Figure~\ref{fig:scatterplots}, which show scatter plots relating both tomographic reconstruction
fluxes and galaxy overdensity to the underlying DM overdensity. At regions of under- and average-density 
($\Delta_\mathrm{dm}\lesssim 1$), the reconstructed \lya\ fluxes are roughly linear with respect to $\Delta_\mathrm{dm}$, 
but the $\Delta_\mathrm{galaxy}$ distribution is nearly flat with respect to $\Delta_\mathrm{dm}$ in this regime. 
The smoothed galaxy distribution only becomes linear with respect to $\Delta_\mathrm{dm}$ at denser regimes, and at $\Delta_\mathrm{dm} \gtrsim 2$ it
exhibits less scatter than the \lya\ forest reconstruction. 
We will pursue these comparisons in more detail in a separate paper \citep[although see, e.g.,][]{kitaura:2012}.

\bfig
\includegraphics[width=0.49\textwidth]{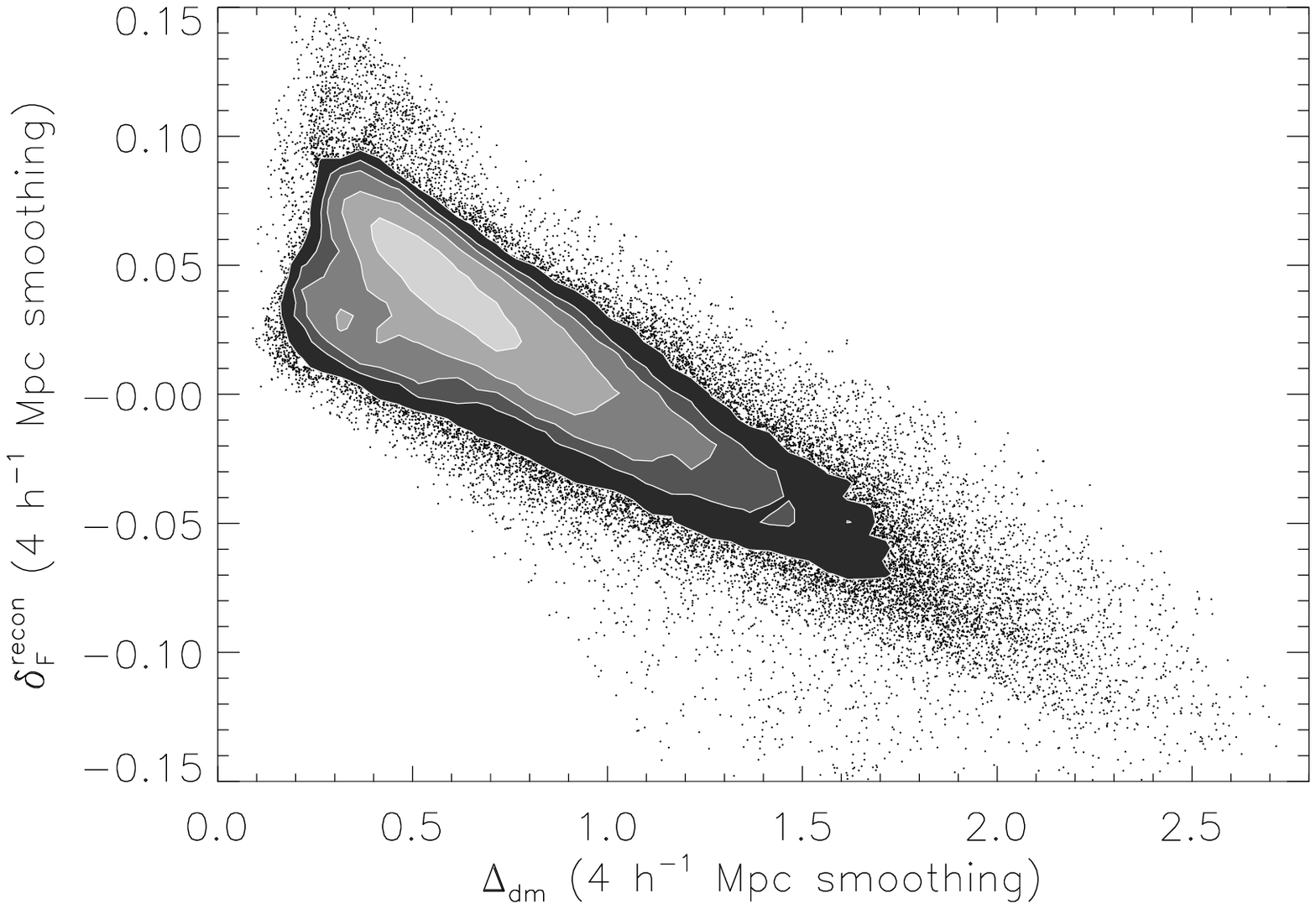}
\includegraphics[width=0.49\textwidth]{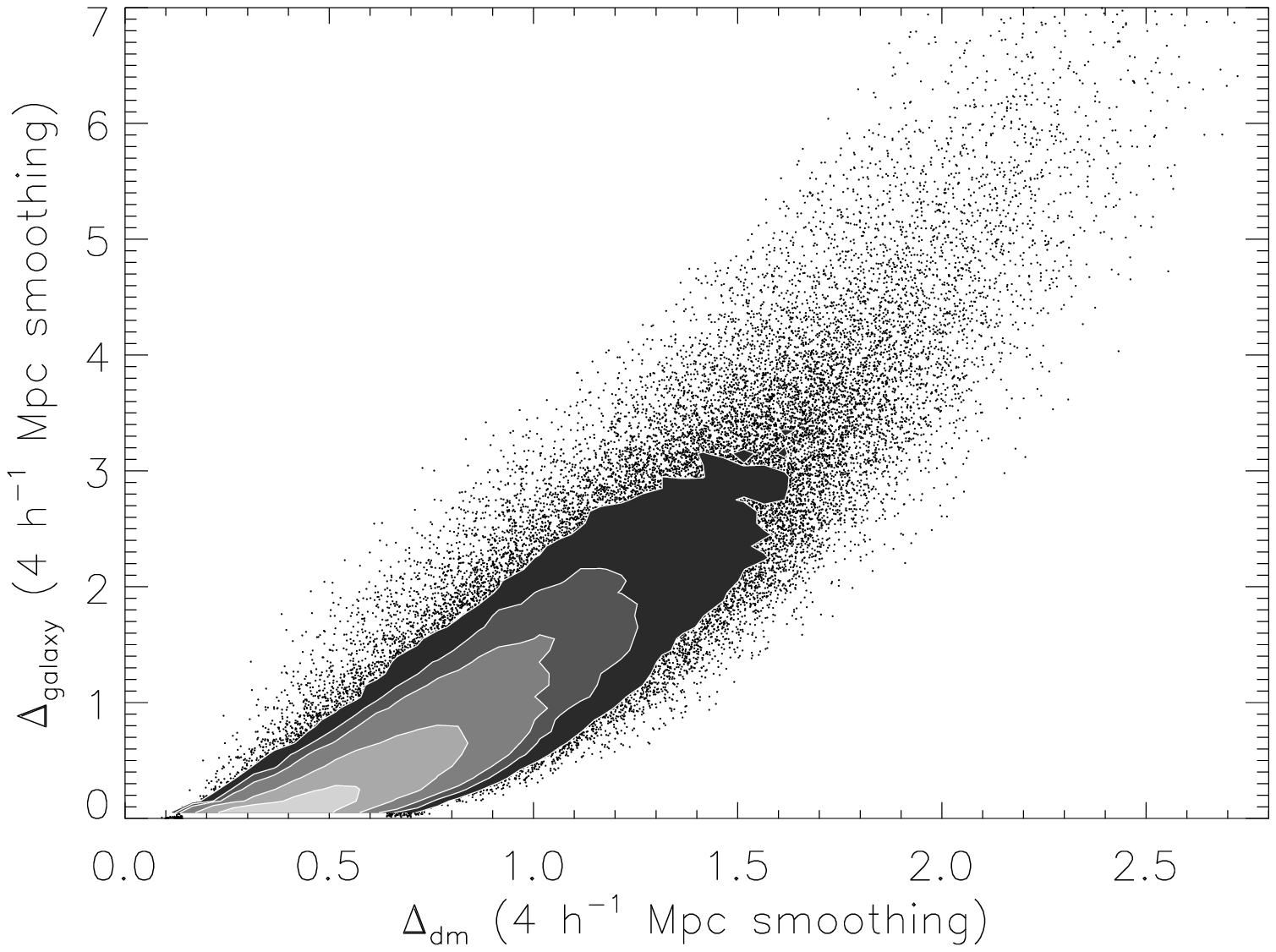}
\caption{\label{fig:scatterplots}
Reconstructed \lya\ forest flux (top) and galaxy overdensity (bottom) from the simulation volume
plotted as a function of the corresponding dark matter overdensity, within $(2\,\mpc)^3$ voxels.
The corresponding visualizations are in the middle panels of Fig.~\ref{fig:recon} and Fig.~\ref{fig:galmap}, 
respectively. The contours denote the 10th, 20th, 30th, 50th and 80th percentiles of the distributions.
The \lya\ forest tomographic reconstruction clearly gives a better mapping of the smoothed DM distribution 
at $\Delta_\mathrm{dm} \lesssim 1$, while the relationship between galaxy overdensity and DM overdensity 
becomes quasi-linear at $\Delta_\mathrm{dm} \gtrsim 1$.
}
\efig

\bfig
\epsscale{1.15}
\plotone{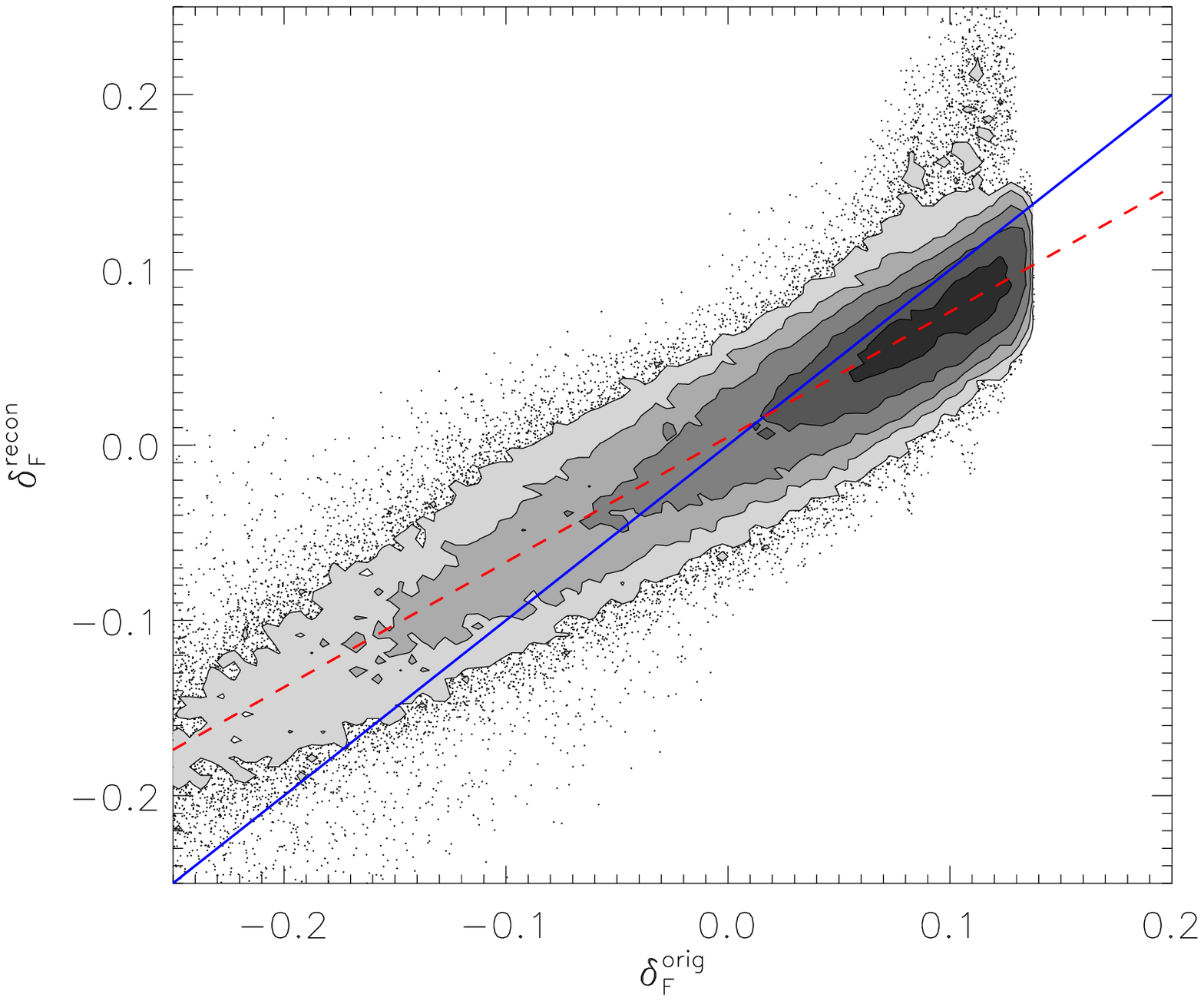}
\caption{\label{fig:recondelta}
Scatter plot of the reconstructed absorption field fluctuations, $\delrecon$, against the true forest fluctuations, 
$\delorig$, in the $\nlos=2218\,\persqdeg$ and $\sigthreed=1.4\,\mpc$ reconstruction as shown in the top-right of Figure~\ref{fig:recon}.
Both fields have been smoothed to $1.4\,\mpc$ with a spherical Gaussian filter.
The contours denote the 5th, 15th, 30th, 50th, and 80th percentiles of the distribution.
The solid blue line is the $\delorig= \delrecon$ relation, while 
the red dashed line shows the best-fit regression linear fit for the points. 
The relationship between $\delrecon$ and $\delorig$ is quite linear with a cross-correlation coefficient of 
$r = 0.93$,
allowing a straightforward correction for the overall bias. 
The distribution of \delrecon\ shown here corresponds to $\snreps = 2.49$.
}
\efig

 \begin{figure*} \centering
\vspace{4em}  
\begin{overpic}[width=0.235\textwidth,unit=1pt]{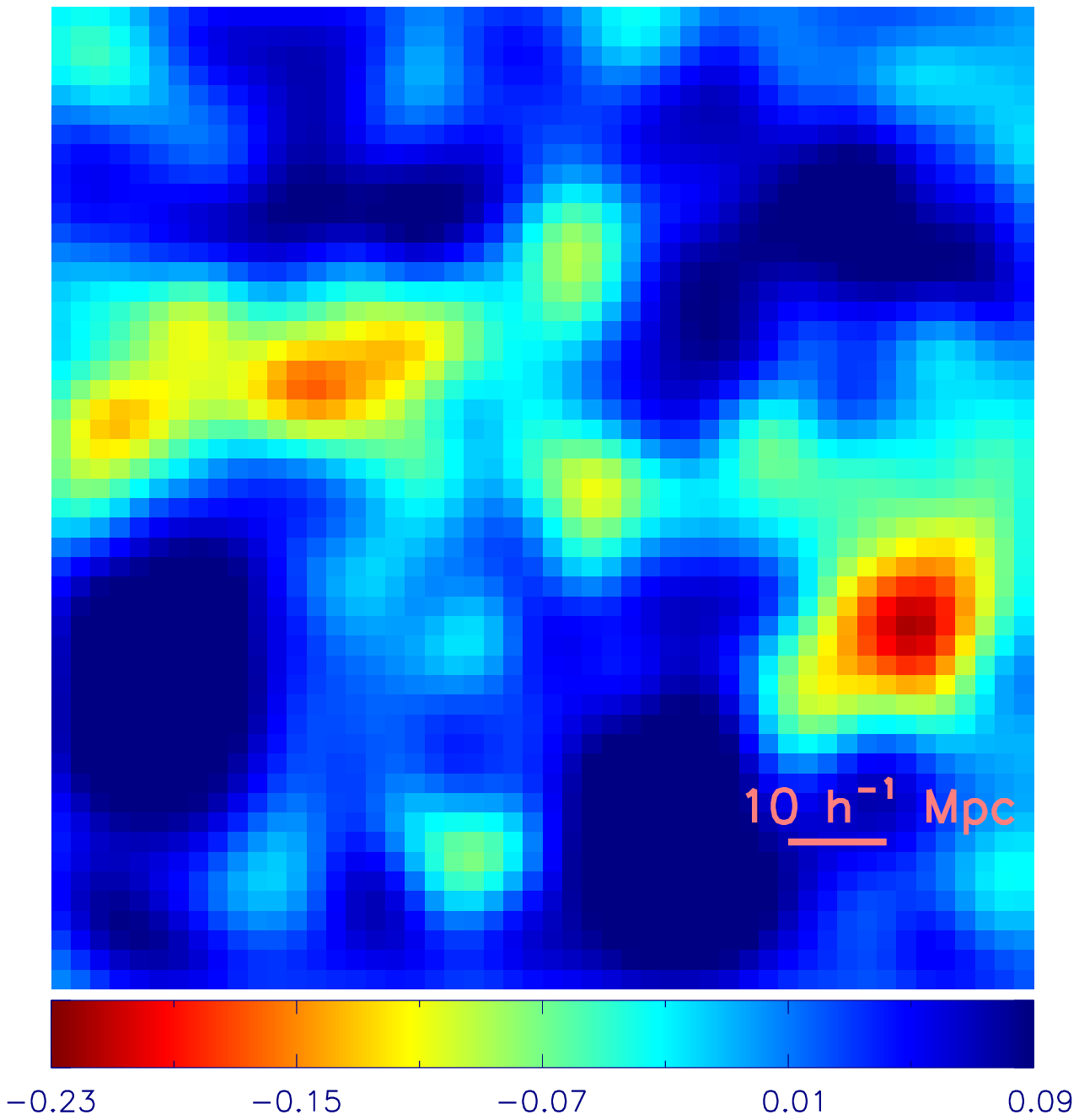} 
\put(16,124){\footnotesize \textsf{True \lya\ Absorption Field}}
\end{overpic} 
\begin{overpic}[width=0.235\textwidth,unit=1pt]{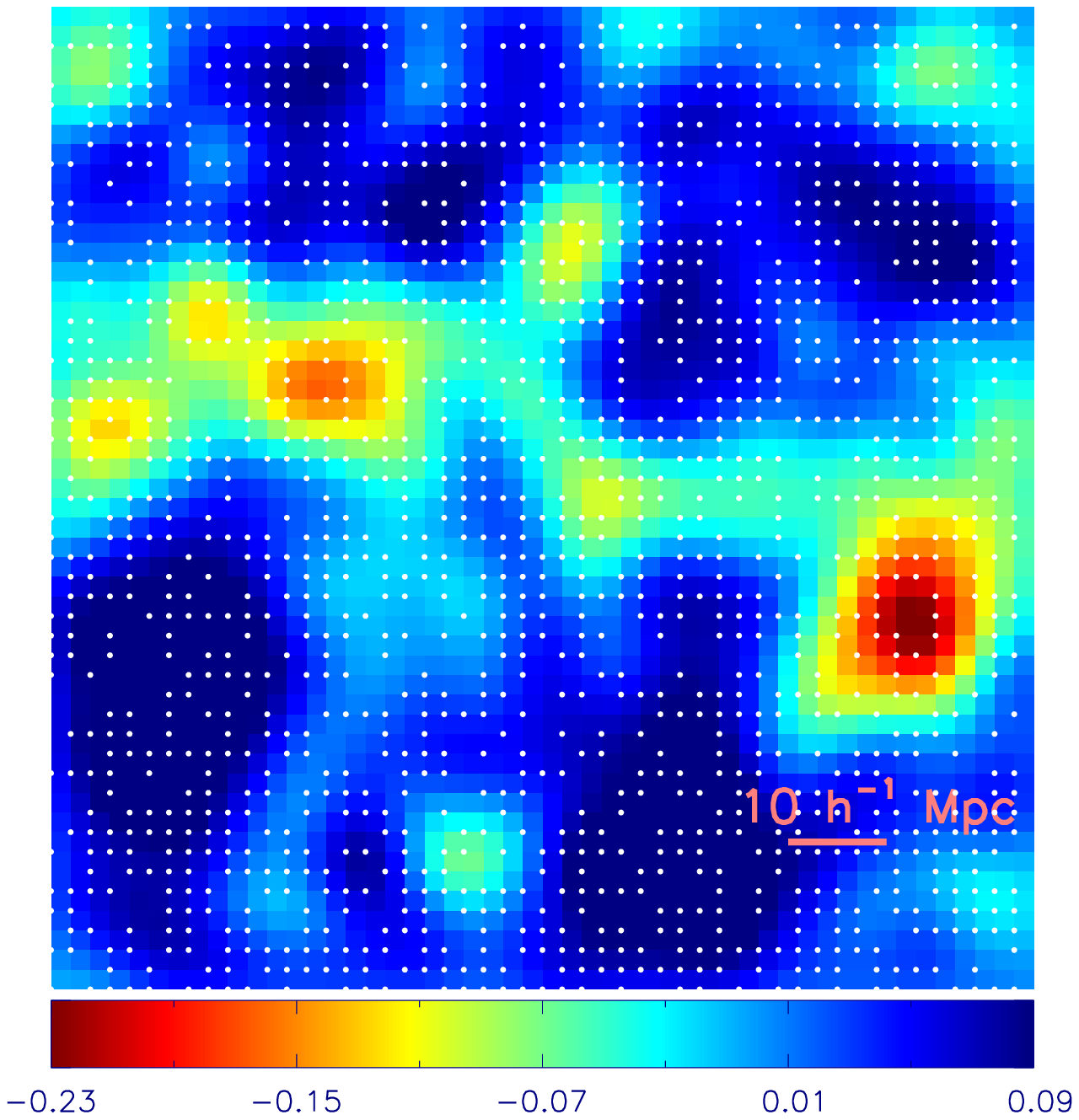} 
\put(5,135){\footnotesize $\mathsf{n_{los} = 971\, deg^{-2},\; t_{exp} = 8\,\hrs}$}
\put(38,124){\footnotesize $\mathsf{SNR_\epsilon = 3.0}$}
\end{overpic}  
\begin{overpic}[width=0.235\textwidth,unit=1pt]{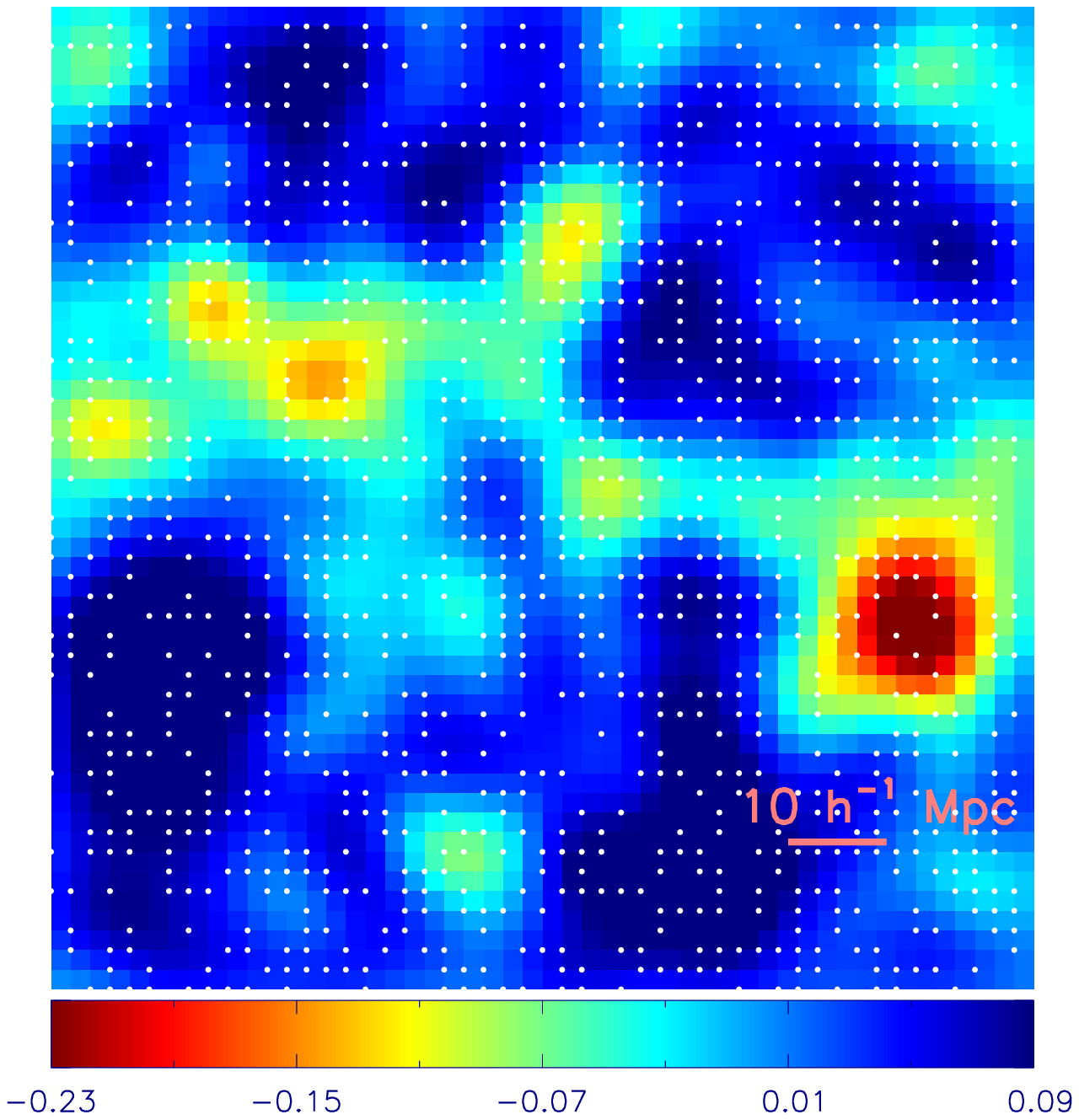} 
\put(5,135){\footnotesize $\mathsf{n_{los} = 657\, deg^{-2},\; t_{exp} = 6\,\hrs}$}
\put(38,124){\footnotesize $\mathsf{SNR_\epsilon = 2.5}$}
\end{overpic}  
\begin{overpic}[width=0.235\textwidth,unit=1pt]{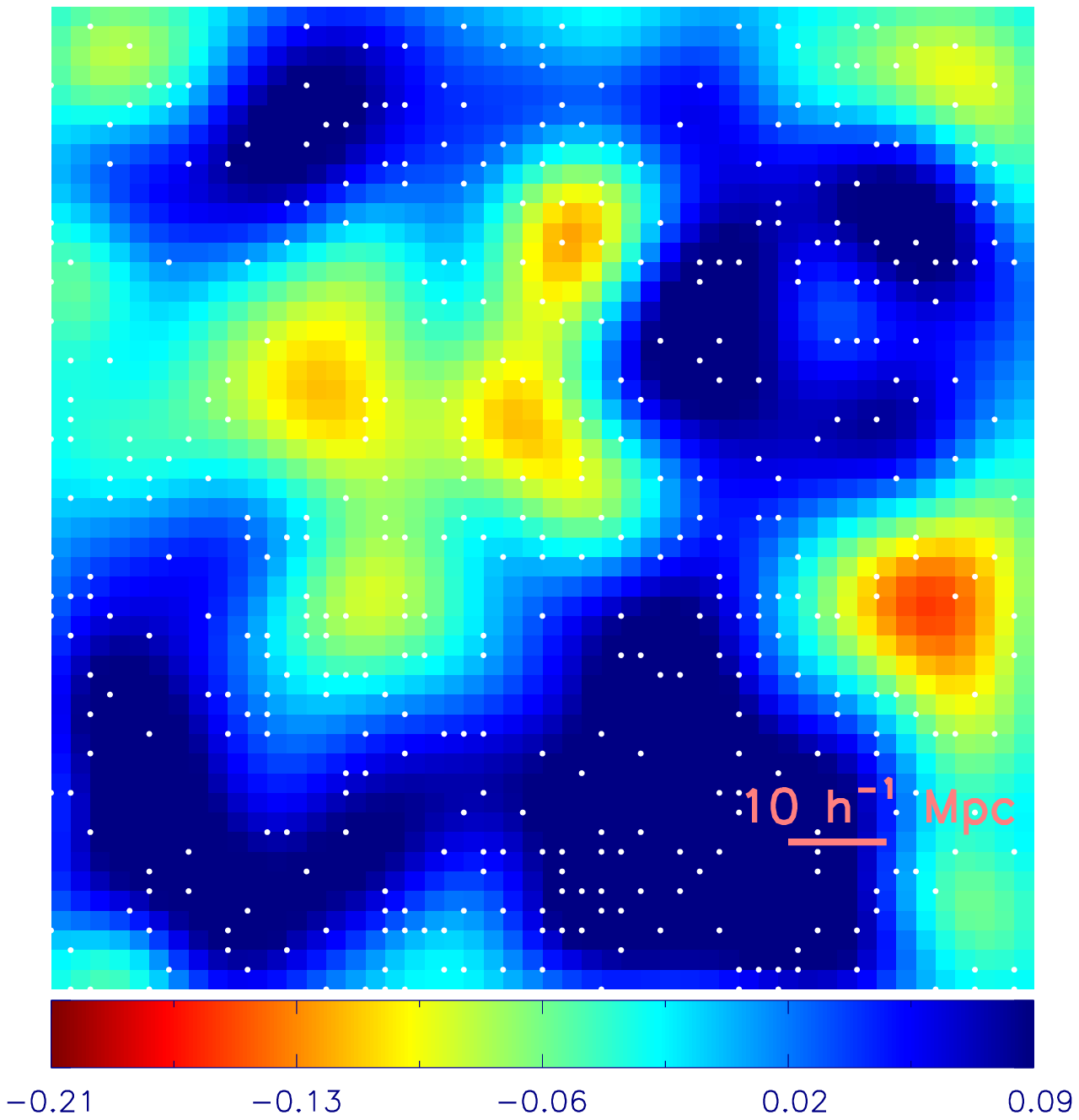} 
\put(5,135){\footnotesize $\mathsf{n_{los} = 112\, deg^{-2},\; t_{exp} = 2\,\hrs}$}
\put(38,124){\footnotesize $\mathsf{SNR_\epsilon = 1.2}$}
\end{overpic}

\caption{\label{fig:snreps}
Simulated tomographic reconstructions of the same field 
generated from different survey parameters $[\texp, \nlos]$ and compared with the true underlying
3D \lya\ forest absorption field (at left). All the maps have been smoothed with a Gaussian kernel of 
$\sigthreed = 3.5\,\mpc$ to facilitate a direct comparison.
The simulation slices have dimensions $(100\,\mpc)^2 \times 2\,\mpc$, the line-of-sight direction is into
the plane of the page. The positions of the \lya\ forest mock skewers that contributed to each reconstruction 
are indicated by the white dots in each map.
The reconstruction signal-to-noise, \snreps, is also labeled and the deterioration of reconstruction fidelity
with lower $\nlos$ and \texp\ is apparent.
}
 \end{figure*} 
 
Apart from visual inspection, we quantify the fidelity of the tomographic maps by
 plotting the scatter of the reconstructed field \delrecon\ against
the full 3D \lya\ forest field \delorig, in which both fields have been binned and smoothed
as described in \S~\ref{sec:wiener}. This is shown in Figure~\ref{fig:recondelta}:
the distribution of points in the scatter plot appears to be quite linear. However, 
we see a bias in the sense that the best-fit linear regression differs from the
 $\delrecon = \delorig$ relation that would be obeyed in the case of a perfect reconstruction. 
This is likely due to the \emph{ad hoc} nature of our Wiener reconstruction algorithm:
 for example, in the final step we smooth the full field with an isotropic 3D Gaussian kernel, even though 
 the geometry of the problem is intrinsically anisotropic between the
 line-of-sight and transverse directions (c.f.\ Equation~\ref{eq:corrfunc}). 
 In this paper we are more concerned with the observational requirements
 of IGM tomographic mapping, so we defer improvements on the algorithm to future work.
 For now, we simply correct $\delrecon$ by the best-fit regression linear function to remove
 the bias; the reconstructed maps shown in Figure~\ref{fig:recon} have already been corrected for this bias.
  
 The $\delrecon$-$\delorig$ comparison also allows us to define a reconstruction signal-to-noise ratio, 
 \snreps, through the signal variance and residual variance:
 \beq \label{eq:snreps_recon}
 \snreps^2 = \frac{\mathrm{Var}(\delorig)}{\mathrm{Var}(\delrecon - \delorig)},
 \eeq
 where $\delrecon$ has been corrected for the bias as described in the previous paragraph.
 For a fixed set of sightlines, \snreps\ can be varied by changing the smoothing scale \sigthreed:
 by smoothing to larger scales on both the reconstruction and true map, the agreement between
 the two is improved and vice-versa.
 
 For a qualitative comparison of how reconstructions at various \snreps\ appear visually, in 
 Figure~\ref{fig:snreps} we show several tomographic maps with different \nlos (and \texp, related through Equation~\ref{eq:texp_nlos}), but 
 smoothed to the same \sigthreed. 
Compared to the true underlying field, the mock data set with $[\texp, \nlos] = [8\,\hrs,\, 971\,\persqdeg]$ gives
a very good reconstruction of the main structures with $\snreps= 3.0$, although there are some discrepancies at  
flux levels close to the $\delta_F = -0.05$, which approximately traces the mean-density of the dark-matter distribution.
The quality of this map is unsurprising, 
as the sightline density corresponds to an average sightline separation of $\dperp = 2.1\,\mpc$
(c.f.\ Equation~\ref{eq:dperp}), 
which oversamples the field at this $\sigthreed = 3.5\,\mpc$ mapping resolution, i.e.\ within a $\sigthreed^2$
patch on the sky, there are $\sim 2.7$ sightlines that sample the area.
Moving down to the survey parameters $[\texp, \nlos] = [6\,\hrs,\, 657\,\persqdeg]$, the resulting value of 
\snreps\ is somewhat decreased to 2.5, but visually the true absorption is still well reconstructed apart from some 
deterioration to delicate features, again at $\delta_F \sim -0.05$. 
This survey's sightline density of $\nlos = 657\,\persqdeg$ corresponds to a sightline separation of $\dperp = 2.6\,\mpc$, 
which is again well-sampled with respect to $\sigthreed=3.5\,\mpc$.
We move on to the right-most panel in Figure~\ref{fig:snreps} and see that the $[\texp, \nlos] = [2\,\hrs,\, 112\,\persqdeg]$
now gives an inferior tomographic reconstruction, with large distortions and none of the delicate filamentary structures reproduced.
This is because the sightline separation, $\dperp = 6.3\,\mpc$, corresponding to this survey depth is too
coarse compared to $\sigthreed = 3.5\,\mpc$, although the agreement can be improved by smoothing to larger \sigthreed.
Qualitatively, $\snreps \approx 2.0-2.5$ seems to be a reasonable threshold for a `good' reconstruction, 
but more quantitative criteria will have to depend on the science goals  
of the survey 
--- these will be explored in future papers where we investigate the utility of IGM tomography for
galaxy environment studies and hunting for the progenitors of galaxy clusters.

 \begin{deluxetable*}{c c c  c c c  c }
\tablecolumns{7}
\tablecaption{\label{tab:epsilon} Mapping Resolution for Different Survey Parameters}
\tablehead{
\multirow{2}{*}{\texp\tablenotemark{a} (hrs)} & \multirow{2}{*}{\glim\tablenotemark{b}} & 
\multirow{2}{*}{$\nlos$\tablenotemark{b} ($\persqdeg$)}  & \multirow{2}{*}{$\dperp$\tablenotemark{d} ($\mpc$)}
& \multicolumn{3}{c}{$\sigthreed$\tablenotemark{d} $(\mpc)$}    \\  \noalign{\vskip 0.3em}
\cline{5-7} \noalign{\vskip 0.3em} 
   &     & & & \colhead{$\snreps=2.0$}     &        \colhead{$\snreps=2.5$}     &  
                                                 \colhead{$\snreps=3.0$}        
                }
                \startdata
2  &  23.6  &  110  &  6.3  &   8.1  &  13.9  &  20.5  \\ 
3  &  23.8  &  230  &  4.4  &   4.6  &   7.2  &  10.5  \\ 
4  &  24.0  &  360  &  3.5  &   3.2  &   4.8  &   6.9  \\ 
5  &  24.1  &  500  &  3.0  &   2.6  &   3.8  &   5.3  \\ 
6  &  24.2  &  660  &  2.6  &   2.3  &   3.2  &   4.5  \\ 
8  &  24.4  &  970  &  2.1  &   1.9  &   2.7  &   3.7  \\ 
10  &  24.5  &  1300  &  1.9  &   1.6  &   2.3  &   3.2  \\ 
12  &  24.6  &  1600  &  1.7  &   1.4  &   2.0  &   2.7  \\ 
16  &  24.8  &  2200  &  1.4  &   1.0  &   1.3  &   1.7 
\enddata
\tablenotetext{a}{Telescope exposure time, assuming the HR-Blue mode on VLT-VIMOS.}
\tablenotetext{b}{Limiting $g$-magnitude of survey, defined by requirement to achieve $\snr\ge 4$ per angstrom at survey limit.}
\tablenotetext{c}{Areal density of sightlines used in reconstruction. Related to \texp\ by Equation~\ref{eq:texp_nlos}. }
\tablenotetext{d}{Typical comoving separation between sightlines in the transverse direction.}
\tablenotetext{e}{Map resolution that would yield the specified reconstruction fidelity \snreps\ for the choice of \texp\ and \nlos.}
\end{deluxetable*}

\begin{figure}
\epsscale{1.25}
\plotone{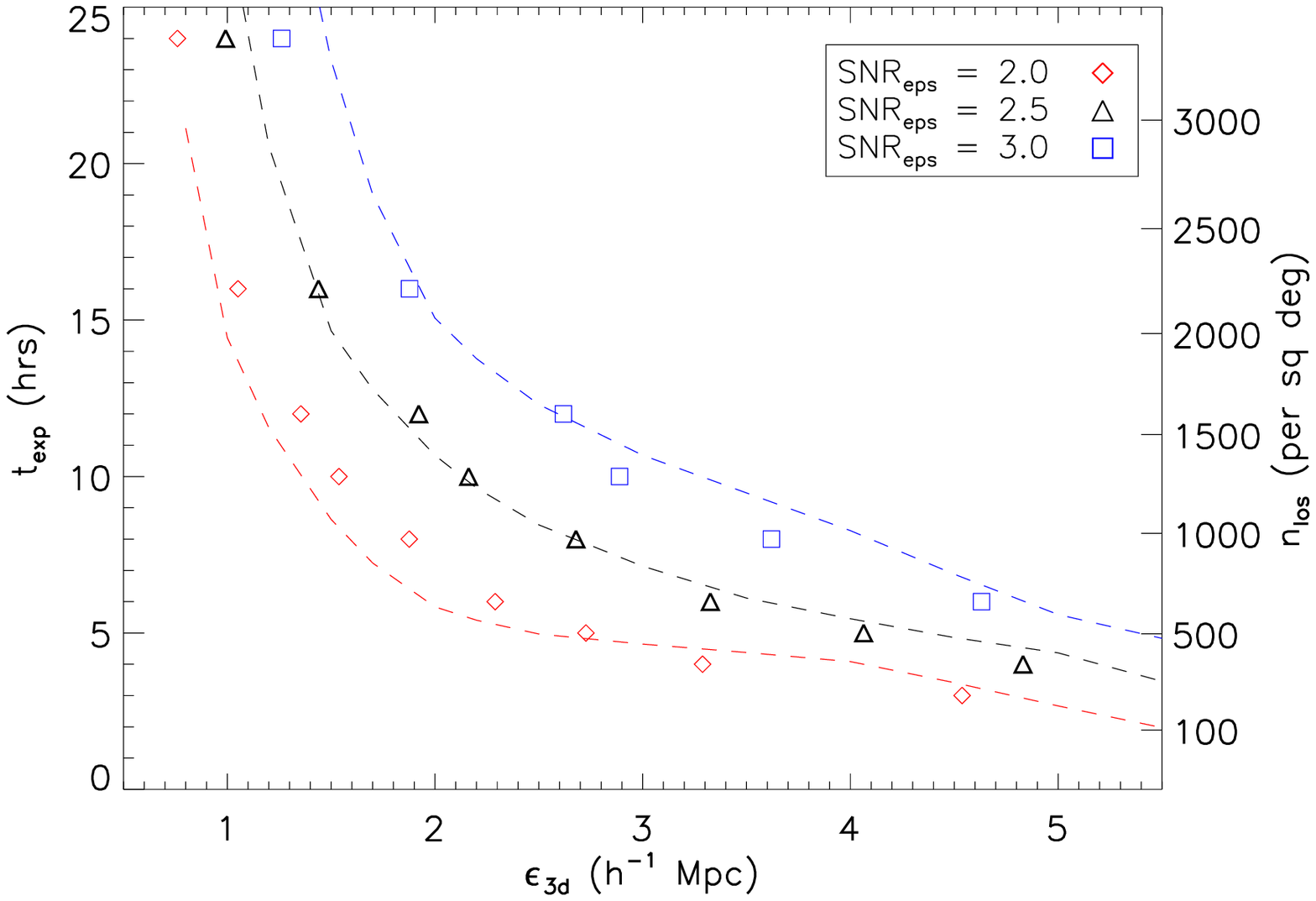}
\caption{\label{fig:texp_eps}
The exposure time, on the VLT-VIMOS spectrograph, required to carry out tomographic reconstructions 
at a given spatial resolution, $\sigthreed$ at various reconstruction signal-to-noise, \snreps (denoted by different colors). 
The symbols are derived from our simulated reconstructions, while 
the dashed-lines show the same quantity estimated with the analytic estimates of \S~\ref{sec:wiener_anal}.
The right-hand axis labels the background source densities achievable given the exposure time, assuming
that the spectra must have at least $\snr \geq 4$ per angstrom.
}
\end{figure}

Intuitively, one expects a reasonable reconstruction when $\sigthreed \gtrsim \dperp$, but there is some leeway
in selecting the smoothing scale \sigthreed\ depending on the desired reconstruction fidelity \snreps.
We can relate the survey parameters $[\texp,\,\nlos]$ as a function of mapping scale
\sigthreed\ at fixed \snreps, by simply varying \sigthreed\ at fixed $[\texp,\,\nlos]$
until the desired \snreps\, is achieved. 
This is shown by the symbols in Figure~\ref{fig:texp_eps}, and also
tabulated in Table~\ref{tab:epsilon}.
At fixed reconstruction fidelity, the \texp\ vs $\sigthreed$ curve remains relatively flat at $\sigthreed \gtrsim 4\,\mpc$, i.e.\
small increases in survey parameters enable large improvements in the mapping resolution. 
For example, going from $\texp=3\,\hrs$ to $\texp=5\,\hrs$ leads to nearly a doubling of the map resolution from
$\sigthreed = 7.2\,\mpc$ to $\sigthreed = 3.8\,\mpc$, at fixed $\snreps \approx 2.5$.
With the current generation of 8-10m telescopes, it therefore makes sense for the first generation of IGM tomographic surveys to 
target the regime where the curves in Figure~\ref{fig:texp_eps} 
begin to rise steeply, i.e.\ 
 $\texp \approx 5-6\,\hrs$ and $\nlos \approx 500-600\,\persqdeg$, that would enable maps with $\sigthreed = 3-4\,\mpc$
 at $\snreps \sim 2.5$. 
To achieve map resolutions of $\sigthreed \lesssim 2\,\mpc$, the survey requirements become increasingly stringent, 
with $\texp \gtrsim 10\,\hrs$ and rising rapidly. 
Tomographic maps with $\sigthreed \approx 1\,\mpc$ resolution would require $\texp \gtrsim 20\,\hrs$ to obtain 
sufficient data from the $\glim \approx 25$ background sources, rendering it very challenging for 
the current generation of telescopes.
However, with future 30m-class telescopes it would require only of order 1-2 hours of integration to obtain the spectra.
Note that this is considerably less than the $\sim 10\,\hrs$ integrations previously assumed to be necessary for 
$\sigthreed \sim 1\,\mpc$ IGM tomography on 30m telescopes.
\citep{steidel:2009, evans:2012}.

\subsection{Analytic Estimates of Reconstruction Noise}\label{sec:wiener_anal}
In this section, we attempt to calculate analytically the relationship between the survey parameters
(\texp, \nlos, \dperp) and the characteristics of the resulting tomographic map ($\sigthreed$, $\snreps$).
The purpose of this calculation is two-fold: (a) to develop intuition for how the various survey parameters
affect the final map, and (b) to provide a `quick-and-dirty' way to determine the 
effect of survey complications such as target-selection inefficiencies, uneven exposure times etc.
As such, we do not attempt to make a rigorous derivation, and will defer such work to a future paper where
we work on improvements to the Wiener reconstruction algorithm.

We begin by defining\footnote{For clarity, in this section we have changed the notation of some variables. The signal $s$
and map $m$ here correspond, respectively, to \delorig\ and \delrecon\ in the previous section.} 
the data as $d = s + n$, i.e.\ the sum of the true underlying signal $s$ and a noise contribution, 
$n$. The signal and noise covariances are denoted as $\mathbf{S} \equiv \langle s \cdot s^{T} \rangle$ 
and $\mathbf{N} \equiv \langle n \cdot n^T \rangle$ respectively, giving the corresponding 
data covariance 
\beq
\mathbf{D} \equiv \langle d \cdot d^T\rangle = \mathbf{S} +\mathbf{N}
\eeq
 if we assume that the noise and signal
are uncorrelated, $\langle s \cdot n^T \rangle = 0$. 
Our reconstructed map is the data convolved by a filter, $\hat{m} = \mathbf{K} \cdot d$, with the covariance 
\beq
\mathbf{\hat{M}} \equiv \langle \hat{m} \cdot \hat{m}^T \rangle = \mathbf{K} \langle d \cdot d^T \rangle \mathbf{K}^T = \mathbf{K} \cdot \mathbf{D} \cdot \mathbf{K}^T,
\eeq
where the hat indicates that the reconstructed map is defined, in principle, on a different set of coordinates 
from the data.
In the case of Wiener filtering, 
\beq
\mathbf{K} = \hat{\mathbf{S}} \cdot (\mathbf{S} + \mathbf{N})^{-1}.
\eeq
In our tomographic reconstructions, both the reconstructed map and true map have been binned on to the same grid, 
so we drop the hat in the notation, e.g.\ $\hat{\mathbf{S}} = \mathbf{S}$.

We can thus write down the covariance of the residual between the reconstructed map and the true underlying field as:
\begin{eqnarray} \label{eq:resid_covar}
\mathbf{R} &\equiv&  \langle (m-s) (m-s)^T \rangle \nonumber \\
&=& ( \mathbf{K} - \mathbf{I} ) \cdot \mathbf{S} \cdot (\mathbf{K}- \mathbf{I})^T + \mathbf{K} \cdot \mathbf{N} \cdot \mathbf{K}^T,
\end{eqnarray}
where $\mathbf{I}$ is the identity matrix and we have used the fact that 
\begin{eqnarray}
m-s &=& \mathbf{K}\cdot (s + n) - s \nonumber \\
&=& (\mathbf{K} - \mathbf{I}) \cdot s + \mathbf{K} \cdot n,
\end{eqnarray}
and again assumed that $\langle s \cdot n^T \rangle = 0$.

We are now in the position to estimate analytically \snreps\ at some map resolution \sigthreed\ 
and telescope exposure time \texp, which sets the noise of the map. In the notation of this section, this is
\beq \label{eq:snreps_anal}
\snreps^2(\sigthreed , \texp) = \frac{\sigma_S^2 (\sigthreed) }{ \sigma_R^2(\sigthreed , \texp)}, 
\eeq
i.e. the ratio of the signal and residual variances from the reconstructions.
In the present case, the numerator, $\sigma_S^2$, is the variance of the 
true \lya\ forest absorption field\footnote{This quantity is more usually denoted $\sigma_F^2$ in the literature} smoothed over standard
deviation \sigthreed. 
Meanwhile, $\sigma^2_R$ is the residual variance from the reconstruction smoothed over
the same scale. 
The latter has a dependence on \texp\ because this determines the pixel noise in the absorption
spectra that go into the reconstructions.

In our case, $\sigma_S^2(\sigthreed)$ is the signal $s$ smoothed over a Gaussian window with 
standard deviation \sigthreed, 
which is easily evaluated by carrying out a volume integral
over the flux power-spectrum:
\begin{eqnarray} \label{eq:sig_variance}
 \sigma^2_S(\sigthreed) & = & \frac{1}{2\pi} \int^\infty_0 k^2 \mathrm{d}k  \int^{1}_{-1} \mathrm{d}\mu \nonumber \\
 & &  \hphantom{howdy} \times P_F(k,\mu)\, W^2(k \sigthreed ),
\end{eqnarray}
where $P_F(k, \mu)$ is the anisotropic 3D \lya\ forest flux power spectrum, $\mu = \cos \theta$ is the ratio between the
parallel component of the wave-vector, $k_\parallel$, and its modulus $k$, and $W(kR)$ is the Fourier space Gaussian filter:
\beq
W(kR) = \exp\left(- \frac{1}{2} k^2 R^2 \right).
\eeq

For the \lya\ forest power spectrum $P_F(k,\mu)$, we use the analytic model \citep[e.g.,][]{mcdonald:2003,mcquinn:2011b,mcquinn:2011} 
\beq \label{eq:pk3dforest}
P_F(k, \mu) = b^2 (1 + \beta \mu^2)^2 P_L(k) \exp(-k^2_\parallel/k^2_D),
\eeq
where $b$ and $\beta$ are the bias and anisotropy parameters respectively \citep{kaiser:1987}, $P_L(k)$ is the linear-theory dark-matter
power spectrum, and $k_D$ parametrizes the small-scale cut-off in the line-of-sight power spectrum from Jeans
and thermal smoothing.
We set $k_D = 0.08\, \mathrm{s}\, \mathrm{km}^{-1}$ as in \citet{mcquinn:2011}, 
but its exact value is unimportant to us as the cutoff is at much smaller scales than we are concerned with.

By analogy with Equation~\ref{eq:sig_variance}, we can evaluate $\sigma_R^2$ as the integral over the
residual power spectrum multiplied by a Gaussian filter.
If we assume translational invariance then all of the matrices in the expression for the residual covariance (Equation~\ref{eq:resid_covar}) 
are diagonal in $k$-space, and we can just replace the matrices with power spectra to obtain an expression for the
residual power spectrum, $P_R(k,\mu)$. In addition, the linear algebra becomes regular algebra.
We then get
\begin{eqnarray} \label{eq:residpower}
P_R &=& (K-1) P_F (K-1) + K (\bar{n}_\mathrm{eff}^{-1} P_\mathrm{los}) K \nonumber \\
&=& \frac{P_F (\bar{n}_\mathrm{eff}^{-1} P_\mathrm{los})^2+ P^2_F\, \bar{n}_\mathrm{eff}^{-1} P_\mathrm{los}}{(P_F + \bar{n}_\mathrm{eff}^{-1} P_\mathrm{los})^2},
\end{eqnarray}
where we have substituted in the Fourier-space Wiener filter $K = P_F / (P_F + \bar{n}_\mathrm{eff}^{-1} P_\mathrm{los})$
in the second line, and do not explicitly write out the dependencies on $k$ and $\mu$.
The term $\bar{n}_\mathrm{eff}^{-1} P_\mathrm{los}$ is an approximation for the \lya\ forest noise power from 
Equation 12 in \citet{mcquinn:2011}, where $P_\mathrm{los}$ is the line-of-sight \lya\ forest power,
\beq \label{eq:neff}
\bar{n}_\mathrm{eff} \equiv \frac{1}{A} \sum_{n}\nu_n, \hphantom{yolo} \nu_n = \frac{P_\mathrm{los}}{P_\mathrm{los} + P_{N,n}},
\eeq
$A$ is the transverse area of the survey, and 
\beq \label{eq:pixelnoise}
P_{N,n} = 0.8 \langle F \rangle^{-2} [(\snr)_n]^{-2} \left(\frac{1+z}{4} \right)^{-3/2}
\eeq
is the line-of-sight power arising from pixel noise in an individual spectrum $n$ that has signal-to-noise 
$(\snr)_n$ (per angstrom), with a redshift dependence that arises from the conversion between observed
wavelength and line-of-sight comoving distance.
For each spectrum $n$ in a mock survey, we can evaluate $\nu_n$ as a function of its spectral signal-to-noise, 
which is in turn set by the source magnitude and exposure time. The sum of all the $\nu_n$ divided by the survey area
gives $\bar{n}_\mathrm{eff}$, which sets the noise in the reconstructions.
\citet{mcquinn:2011} assumed a fixed value of $P_\mathrm{los}$ since the line-of-sight power
spectrum is approximately white at the scales they were concerned with, but we carry out the full
evaluation of $P_\mathrm{los}$ as a function of $k_\parallel = k\,\mu$.

With the residual power spectrum in hand (Equation~\ref{eq:residpower}), we then compute the 
residual variance within a smoothing window \sigthreed\ in an analogous integral to Equation~\ref{eq:sig_variance}.
This now allows us to compute, through Equation~\ref{eq:snreps_anal}, 
the $\snreps$ expected for a given combination of \sigthreed\ and
\texp. The analytic curves for \texp\ against \sigthreed, at fixed \snreps, is plotted as the dashed-lines
in Figure~\ref{fig:texp_eps}. The only free parameters we have in this model is $b$ and $\beta$ in the
3D \lya\ forest power spectrum (Equation~\ref{eq:pk3dforest}) --- the curves in Figure~\ref{fig:texp_eps}
assume $b = 0.21$ and $\beta = 0.5$, which are both consistent with the fits of \citet{slosar:2011}.
The analytic curves have the same qualitative behavior as the points estimated from the simulated reconstructions, although 
 the simulation points also have an error from the Poisson sampling of background sightlines.
Note that in this analytic calculation, we have assumed that the Wiener filter uses the correct \lya\ forest
correlation function $\mathbf{S}$, whereas in the simulations we have used the \emph{ad hoc} form 
in Equation~\ref{eq:corrfunc} as suggested by \citet{caucci:2008}. This could also explain part of the
discrepancy seen in Figure~\ref{fig:texp_eps}.
Nevertheless, this analytic approach works surprising well and provides us with some intuition as to how the 
tomographic reconstructions scale with the quality and quantity of survey data.

\bfig
\epsscale{1.25}
\plotone{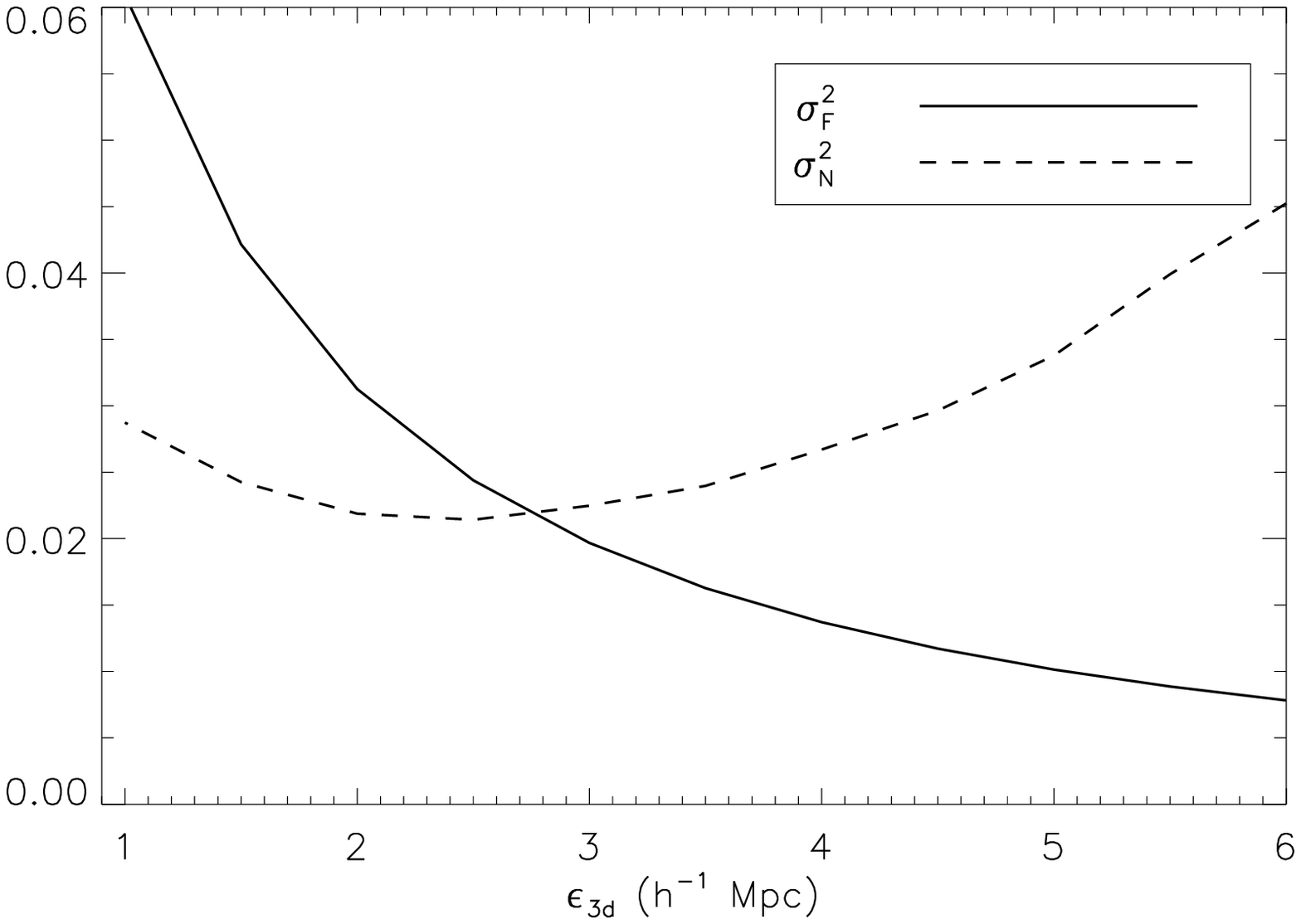}
\caption{\label{fig:noise_vs_f}
The \lya\ forest flux variance  $\sigma^2_\mathrm{F}$ and shot-noise variance $\sigma^2_\mathrm{N}$, 
as a function of reconstruction scale \sigthreed. In this particular case we have assumed that $\sigthreed = \dperp$, 
and set $\bar{n}_\mathrm{eff} = \nlos$ to remove the contribution from pixel noise in the spectra and show only the shot-noise
from the random sampling of sightlines.
}
\efig

An interesting insight from this analysis is that in the hypothetical
case of infinite spectral signal-to-noise, $\bar{n}_\mathrm{eff}
\rightarrow \nlos$ such that the noise term in the residual power
spectrum does not go to zero. In other words, for a finite set of
skewers there is always a shot-noise contribution from the random
sampling of sightlines even in the absence of pixel noise in the
spectra.  
We find that as \texp\ goes to infinity, \snreps\ asymptotes
to a value that increases with \nlos\ or, equivalently, decreases with
\dperp.  For example, if we set $\sigthreed = \dperp$, then we find
$\mathrm{SNR}_{\epsilon,\infty} = [2.56, 2.17, 1.88]$ for $\sigthreed
= [2, 3, 5]\,\mpc$.  The reason for this is that the \lya\ forest
variance, $\sigma^2_F(\sigthreed)$ declines with \sigthreed\ (dashed
line in Figure~\ref{fig:noise_vs_f}), but the variance from the shot
noise term,
 \begin{eqnarray} \label{eq:shotnoise_variance}
 \sigma^2_N(\sigthreed) & = & \frac{1}{2\pi} \int^\infty_0 k^2 \mathrm{d}k  \int^{1}_{-1} \mathrm{d}\mu \nonumber \\
 & &  \hphantom{howdy} \times \nlos^{-1} P_\mathrm{los}(k,\mu)\, W^2(k \sigthreed ),
\end{eqnarray}
remains roughly constant if we assume $\sigthreed = \dperp$.
 This is because as \nlos\ is decreased, 
\sigthreed\ increases and hence
$P_\mathrm{los}$ is also smoothed over a larger window which reduces its value
 and keeps the ratio of both quantities roughly constant.
 This shot-noise variance $\sigma^2_N$ is shown by the solid curve 
 in Figure~\ref{fig:noise_vs_f}.
 At $\sigthreed \gtrsim 3\,\mpc$, 
the shot-noise variance dominates over the flux variance.
This behaviour is reflected by the fact that to achieve reconstructions at fixed \snreps, the ratio $\sigthreed/\dperp$ 
decreases with $\dperp$ as can be seen in Table~\ref{tab:epsilon}. 

So far in this paper, we have tied the availability of background sightlines, \nlos, to the exposure time, \texp, 
based on the assumption that a minimum signal-to-noise must be achieved on the source continuum to obtain a source
redshift. This is a conservative assumption, as $\sim 50\%$ of LBGs have \lya\ emission lines \citep{shapley:2003} 
that enable a redshift measurement at fainter magnitudes. 
Although these fainter spectra will have noisy \lya\ forest pixels, 
they will reduce the shot-noise in the survey and allow significant improvements in the 
tomographic reconstruction, whether
in terms of $\sigthreed$ or $\snreps$.
With our analytic derivation in hand, we are in a position to rapidly explore more complicated survey scenarios.
For example, we find that a survey with $[\texp, \nlos, \glim] = [4\,\hrs,\, 360\,\persqdeg,\, 24.0]$,
which obeys the $\snr \ge 4$ per angstrom
requirement, enables
a $\sigthreed = 5\,\mpc$ reconstructed map with $\snreps = 2.38$. 
If we keep \texp\ fixed but observe $\approx 1000$ additional faint targets per square degree down to $\glim = 24.5$, 
and assume a
50\% success rate in measuring the redshift, this gives $\sim 500$ additional sightlines that contribute to the reconstruction. 
This improves the reconstruction fidelity, which is now
increased to $\snreps = 2.73$ at the same $\sigthreed = 5\,\mpc$. 
Alternatively, one might choose to smooth the map to a smaller scale, 
$\sigthreed = 4.2\,\mpc$, to retain the same reconstruction fidelity $\snreps = 2.38$ as the fiducial survey.

This strategy only works at $\sigthreed \gtrsim 3\,\mpc$ where the shot-noise contribution is dominant over the intrinsic
forest power. At smaller scales ($\sigthreed \lesssim 2\,\mpc$), there is little gained from adding additional sightlines, and 
the only way to improve $\snreps$ is to increase the integration times.

\section{Discussion} \label{sec:discussion}
Beyond what we have presented in this paper, there are questions regarding \lya\ forest tomography
that we have not discussed. 

An important issue that needs to be addressed is that of continuum-fitting in the LBG spectra. 
At the low signal-to-noise ratios that comprise the typical spectra in a tomographic survey sample
(Figure~\ref{fig:noisy_spectra}), it would not
be possible to directly estimate the continuum from regions in the \lya\ forest deemed free from absorption
\citep[e.g.,][]{savaglio:2002}. 
However, while there are likely to be inhomogeneities in the intrinsic LBG spectrum between restframe \lya\ and 
Ly$\beta$ wavelengths, these occur mostly from stellar/interstellar medium absorption lines that can be masked
once the LBG redshift is known --- \citet{shapley:2003} have shown that these are relatively limited in number
within the \lya\ forest wavelength range of the average LBG at those redshifts.
The remaining spectral undulations appear to be of order $\lesssim 20\%$ RMS in star-forming
galaxies with sub-solar metallicities \citep[c.f.][]{leitherer:2011,heckman:2011}.
These can be treated using the various PCA techniques that have been developed to estimate 
quasar continua \citep{suzuki:2005, paris:2011, lee:2012a} from $\lambda > 1216\,\ang$.
Indeed, these techniques are likely to work better on LBGs than quasars: the spectra on either side of the
intrinsic galaxy \lya\ wavelength are physically correlated, whereas in the case of quasars
 there is little correlation between the continuum-slopes either side of $\lambda \sim 1200\,\ang$ 
\citep{telfer:2002}, the physical origin of which is still a mystery.
In any case, we have shown in this paper that the signal-to-noise requirements for a tomographic survey are quite
modest, so the continuum estimation do not have to perform much better than the $\sim 15-25\%$ pixel noise
RMS at the survey limiting magnitude.
However, the need for continuum estimation means that a reasonable portion of the LBG spectra at $\lambda > 1216\,\ang$
needs to be observed in order to identify the intrinsic absorption lines in a region free of \lya\ forest absorption.

The tomographic reconstruction algorithm also could benefit from further
optimization. Although the Wiener interpolation-based scheme we have used here seems to work 
reasonably well, there are various \emph{ad hoc} aspects that could be improved upon. 
For example, the correlation function (Equation~\ref{eq:corrfunc}) that relates the data cells to the mapping volume should 
in principle be derived from recent measurements of the 3D \lya\ forest correlation function \citep[e.g.,][]{slosar:2011}. 
The algorithm could also be improved by incorporating adaptive and/or anisotropic smoothing as the final step, instead of 
the isotropic Gaussian smoothing we are currently implementing.
Moreover, techniques need to be refined to deconvolve the dark-matter overdensity field and peculiar
velocity fields from the 3D absorption maps, e.g. along the vein of \citet{nusser:1999} and \citet{pichon:2001}.
These techniques will need to take into account the uncertainties in the IGM thermal parameters
such as the temperature at mean density, temperature-density relationship and Jeans' scale.

We have also not touched on the practical issue of target selection: the background 
LBGs and quasars need to be identified as such from imaging data in order to be targeted for
spectroscopy. The selection of $z\sim 2$ LBGs through
the BX criterion is relatively efficient ($\sim 85\%$ at $\mathcal{R} \gtrsim 23.5$), but this is a relatively
broad-brush selection that captures LBGs in the range $1.7 < z < 2.6$ \citep{steidel:2004}. For a tomographic
survey it is desirable to have a fine-tuned target selection to identify the sources that maximize \lya\ forest
coverage at the redshifts of interest. 
It would therefore be useful to apply sophisticated density estimation techniques 
\citep[e.g.,][]{richards:2004,bovy:2011} on training sets of known LBGs
to reduce the contamination rate. 
Quasar target selection has evolved considerably in recent years, and with color-color selection alone the 
success rate is $\sim 60\%$ \citep[e.g.][]{ross:2012,bovy:2011}, 
although this increases significantly if UV or IR data is included \citep{bovy:2012}. 
Since the first generation of tomographic surveys are likely to be in well-studied fields such as COSMOS
\citep{scoville:2007} or GOODS \citep{dickinson:2003} that have ample multi-wavelength data, 
we expect target selection to be straightforward, and in any case the spectrographs (see \S~\ref{sec:instruments})
will have target densities of order several thousand per square degree, affording a relatively inefficient ($\lesssim 50-60\%$) target selection
assuming $\nlos \lesssim 1000 \,\persqdeg$ --- this is the success rate of the \citet{steidel:2004} sample, which used only
3-color selection.

\subsection{Potential Science Applications}
Due to the hitherto distant prospects for mapping large-scale structure at $z \sim 2$, 
there has been a dearth of literature on its scientific possibilities.
Here, we outline some possible applications for such maps. Some of these themes will be explored in more 
detail in subsequent papers, and we also invite others to contribute their ideas.

\textbf{Galaxy Environments:} The $z\sim 2-3$ epoch is a particularly
  interesting time for the study of galaxy formation, as the
  star-formation rate density of the Universe peaked at this epoch,
  whilst the present-day Hubble sequence of galaxies were still in the
  process of being assembled.  However, while the galaxies in this
  epoch have been intensively studied both through imaging and
  spectroscopy, it has been extremely challenging to study them as a
  function of environment \citep[although see][]{diener:2013}.
  \lya\ forest tomographic maps on scales of $\sigthreed \sim 1-4
  \,\mpc$ would reveal the impact of environment on the various galaxy
  properties at $z\sim 2$, e.g.\ color, morphology, star-formation
  rate, gas properties. Indeed, since the \lya\ forest absorption is a
  continuous tracer of the underlying dark matter field at
  overdensities of $\delta \rho / \langle \rho \rangle \sim 1$, it
  would be a less biased tracer of the density field than even the
  most comprehensive galaxy redshift surveys (Figure~\ref{fig:recon}). 
  However, there are few
  papers in the literature that directly study galaxy properties as a
  function of the underlying DM field (as opposed to using groups or
  clusters), so some work needs to be done to tie theories of galaxy
  formation directly to the \lya\ forest field.  Moreover, the
  $\sigthreed \sim 1\,\mpc$ scale of the most ambitious tomographic
  reconstructions (e.g.\ right panel of Figure~\ref{fig:recon}) corresponds to $\sim 400\,\mathrm{kpc}$ proper distance at
  those redshifts. This approaches the `circumgalactic medium' (CGM)
  scale for a typical LBG, and would provide valuable insights into
  the gaseous \ion{H}{1} environment of such galaxies. However, since
  the size of the CGM scales with virial radius, coarser-resolution
  tomographic maps with $\sigthreed$ of several Mpc could already
  probe the CGM of galaxy groups and protoclusters.

\textbf{Galaxy Protoclusters:} Even though massive galaxy clusters have
  been discovered out to $z \gtrsim 1.5$ \citep[e.g.,][]{nastasi:2011,gobat:2011,brodwin:2012,muzzin:2013}, the mechanisms for their
  formation is not well-understood due to the difficulty in mapping
  large-scale structure at $z \gtrsim 1$.  In the hierarchical view of
  large-scale structure, galaxy clusters are believed to form through
  the mergers of smaller protoclusters at the intersections of the
  filaments in the cosmic web. With large-scale structure maps from
  \lya\ forest tomography, it should thus be possible to directly
  identify cluster progenitors.  It is likely that this should be
  feasible through coarser maps with resolutions of several $\mpc$; in
  an upcoming paper we will investigate methods to identify
  protoclusters through IGM tomographic maps.
  Selecting galaxy protoclusters directly through through their signature in the
  large-scale structure would be more systematic and thorough than 
  either serendipitous discoveries \citep[e.g.,][]{steidel:2005} or
  by targeting galaxy overdensities around high-redshift radio galaxies 
  \citep{overzier:2006,venemans:2007,hatch:2011}.

 
\textbf{Topology:} The topology of large-scale structure has yet to be
   studied beyond the Local Universe.  \citet{caucci:2008} have
   already shown that \lya\ forest tomographic maps can effectively
   recover the large-scale topology of the Universe at $z \sim 2$.
   The measurement of cosmic topology at $z \sim 2$ will directly test
   the idea that large overdensities such as galaxy groups and
   clusters form at the intersection of filaments 
   in the cosmic web \citep[see, e.g.,][]{bond:1996,kravtsov:2012}.  With the
   higher-resolution maps $\sigthreed \approx 1\,\mpc$, it could also
   be possible to directly test the cold-flow accretion picture of
   galaxy formation \citep{keres:2005, keres:2009, dekel:2009}, 
   in which star
   formation in galaxies is fed by cold gas and dwarf galaxies that
   stream in along the filaments in the cosmic web.  While this mode
   of accretion is challenging to constrain through traditional 1-dimensional
   absorption-line studies due to the modest covering-factor of cold streams
   \citep{faucher-giguere:2011,fumagalli:2011,fumagalli:2014,hennawi:2013}, 
   and the challenge of finding background quasar sightlines at small
   impact parameter to foreground galaxies \citep{crighton:2011, rudie:2012}
a tomographic map of the IGM with
   $\sim 1\,\mpc$ resolution will be a powerful test of this scenario.
 Another interesting possibility is to use the characteristic topology of inflationary cold dark matter cosmology as a 
 standard ruler to measure the expansion rate of the Universe \citep{park:2010, zunckel:2011}. 
 This will require mapping out large volumes, but the scales necessary for the measurement are so 
 large ($\sigthreed \gtrsim 15\,\mpc$) that this could be feasible 
 through the main surveys of MS-DESI (see below) or even BOSS. 
 
\textbf{Power Spectrum and Cross-Correlations:} Beyond tomographic maps, the highly dense sets of \lya\ forest
 spectra obtained for such surveys will also be powerful for \lya\ forest auto-correlation and cross-correlation studies.
 The BOSS \lya\ Forest Survey \citep{dawson:2013,lee:2013} has pioneered the measurement of the 3D \lya\ forest auto-correlation on scales of
 $\Delta r \gtrsim 10\,\mpc$ \citep{slosar:2011, busca:2013, slosar:2013}, but on smaller ($\Delta r \lesssim 10\,\mpc$) scales the 
 number of available pixel pairs is limited due to the comparatively large transverse separations ($\sim 15\arcmin$)
 between BOSS quasars. A tomographic
 survey that targets $\sim 2000$ spectra over 1 square degree would give twice the number of $\Delta r \lesssim 10\,\mpc$ 
  pixel pairs compared with the final $\sim 160000$ quasar BOSS \lya\ forest sample. 
The measurement of the small-scale \lya\
  forest 3D auto-correlation (or, equivalently, the 3D power spectrum) would place constraints on the underlying dark-matter
  power spectrum at that epoch, as well as the gas temperature of the IGM (A.\ Ari\~no, in prep).
  Beyond the auto-correlation, the \lya\ forest has proven its utility in cross-correlation with other large-scale structure tracers, 
  such as with damped \lya\ systems \citep[DLAs,][]{font-ribera:2012}, and quasars \citep{font-ribera:2013}.
   A dense grid of absorption sightlines would enable cross-correlation studies with e.g., weak-lensing maps 
   \citep{massey:2007}, 
   cosmic microwave background lensing convergence maps \citep[e.g.,][]{sherwin:2012}, 
  and the cosmic near-infrared background \citep[e.g.,][]{fernandez:2010}.

\subsection{Spectrographs and Telescopes for \lya\ Tomography} \label{sec:instruments}
We have established in this paper that surprisingly noisy moderate resolution spectra are sufficient to carry out
IGM tomographic reconstructions from \lya\ forest absorption spectra, in many cases requiring exposure times that are
already feasible on the current generation of 8-10m telescopes.  Here, we outline several regimes in which \lya\ forest
tomography can be carried out on existing and near-future instruments that vary in aperture, field-of-view (FOV) and multiplexing.

\textbf{Medium-FOV Spectrographs on 8m Telescopes:} Throughout this paper, we have used the exposure time calculator for the VIMOS spectrograph \citep{le-fevre:2003} on the 8m VLT to 
predict exposure times. As we have seen in Table~\ref{tab:epsilon}, $\texp = 5\,\mathrm{hrs}$ will yield sufficient signal-to-noise  
to identify complete samples of $g = 24.1$ LBGs at $\zbg \sim 2-3$. 
This gives sufficient source density to create tomographic maps with $\sigthreed \approx 3-4\,\mpc$. With the $4\times7\arcmin \times8\arcmin$ field-of-view on 
VIMOS, it would take about 18 pointings to cover an area of $1\, \mathrm{deg}^2$, or about 140 hrs including overheads. 
Assuming a line-of-sight distance of $\sim 250\,\mpc$ is covered at the same minimum \nlos\ (requiring a total projected source 
density of $\approx 1.8\,\nlos$; see \S~\ref{sec:nlos}), this corresponds to a comoving volume of $\approx 1.1\times 10^6 \mpccube$.
This is equivalent to the volumes covered by the VVDS \citep{marinoni:2008} and zCOSMOS \citep{kovac:2010} 
galaxy redshift maps out to $z \sim 1$, 
but with a considerably less elongated geometry that will allow the characterization of structures spanning the transverse direction
to the line-of-sight. The LRIS spectrograph \citep{oke:1995} on the 10m Keck telescope has a field-of-view
 $\onequarter$ that of VIMOS, but its larger aperture, better blue throughput and lower observing overheads mean that it would take the same overall time to carry out the $1\,\mathrm{deg^2}$ survey.
 
\textbf{Wide-FOV Spectrographs on 8m Telescopes:} The Subaru Prime-Focus
  Spectrograph \citep{takada:2014} is a $1\,\mathrm{deg}^2$
  field-of-view spectrograph with
  2400 fibers currently planned for the 8.2m Subaru telescope. While
  the spectrograph throughput and telescope aperture is similar to
  VIMOS, it has the advantage of a much larger ($\sim 18\times$)
  field-of-view.  The $16\,\sqdeg$ galaxy evolution survey described
  in \citet{takada:2014} involves obtaining redshifts for $\sim 1800$
  LBGs per square degree at $2 < z < 3$, with $\texp = 3\,\hrs$
  exposures. This can be used to generate an IGM tomographic map over
  a comoving volume of $\approx 1.8\times 10^7\,\mpccube$ covering
  $2.1 \leq z \leq 2.4$ --- our analytic estimates indicate that a map
  with $\sigthreed = 3.5\,\mpc$ (similar to the VIMOS/LRIS map
  described above) would have a reconstruction fidelity of $\snreps
  \approx 2.4$, i.e.\ with a quality similar to the 2nd panel from the right in Figure~\ref{fig:snreps}.  
Another possibility is to use the spectrograph for
  dedicated deep observations with $\texp \sim 10$hrs of $\glim
  \approx 24.5$ sources that will reach the sightline densities
  sufficient for $\sigthreed = 1.5\,\mpc$ tomographic maps.  These
  scales, corresponding to $\approx 400\,\mathrm{kpc}$ physical at
  $z\sim 2.3$, is close the regime of interest for circumgalactic
  medium (CGM) studies, and will enable great insights into the
  process of galaxy formation.

\textbf{Wide-FOV Spectrographs on 4m Telescopes:}
The Mid-Scale Dark Energy Spectroscopic Instrument \citep[MS-DESI,][]{levi:2013} is a 5000-fiber spectrograph with a $7\,\mathrm{deg}^2$
field-of-view that is intended to be mounted on the 4m Mayall telescope at the
Kitt Peak National Observatory. The main survey will target LRGs and ELGs out
to $z \sim 1.4$, and \lya\ forest quasars at $z \sim 2.3$ over 14,000 $\mathrm{deg}^2$ to measure the BAO signal. The main 
\lya\ forest survey will target $g \lesssim 23$ quasars at a projected area density of $\sim 50\,\persqdeg$
 that is too low for tomographic reconstruction on scales we are 
interested in. However, the instrument may be made available for community use, so we can envision a dedicated survey
for \lya\ tomography. Such a survey could aim for $\nlos \approx 200\,\persqdeg$ to a magnitude limit of $\glim \approx 23.8$
to obtain maps with $\sigthreed \approx 7\,\mpc$. This would
require $\texp \approx 12\,\mathrm{hrs}$ integrations per pointing 
(we have simply rescaled the VIMOS \texp\ values to a 4m aperture).
Since each pointing would cover a volume of $\approx 8\times 10^6\,\mpc$, 
despite the long exposure times it will be possible to build up large volumes rapidly
to, e.g., identify large numbers of galaxy protoclusters or precisely measure
the topology of large-scale structure.

\section{Conclusions}
In this paper, we have conducted a detailed study of the observational
requirements necessary to carry out IGM tomography, at
various map resolutions, by Wiener interpolation over dense grids of
\lya\ forest absorption spectra.  Using empirical luminosity functions for
background quasars and LBGs, the transverse separation between
sightlines is $\dperp = [1, 2, 3, 5]\,\mpc$ at limiting magnitudes of
$\glim \approx [25.0, 24.5, 24.1, 23.7]$.  With the \emph{ansatz} that
the source separation roughly sets the resolution $\sigthreed$ of the
reconstructed map, this argues that tomographic reconstructions on
scales of $\sigthreed = 3-5\,\mpc$ is feasible with moderate-resolution ($R \sim 1000$) spectra
of $\glim \approx 24$
sources, which is accessible to the current generation of 8-10m
telescopes and instrumentation.

We directly tested this with Wiener reconstructions of mock \lya\ forest absorption spectra generated from numerical
simulations, in which we have added realistic pixel noise based on the source luminosity functions and assumed telescope 
exposure times, \texp. Assuming a conservative signal-to-noise requirement ($\snr \ge 4$ per angstrom), necessary to 
measure the source redshifts, exposure times of $\texp = [3,5,8,16]\,\hrs$ on an 8m telescope can obtain background sightlines 
at sufficient source densities to create good-quality ($\snreps = 2.5$) tomographic maps 
with spatial resolutions of 
$\sigthreed = [7.2, 3.8, 2.7, 1.4]\,\mpc$ (e.g., Figure~\ref{fig:recon}). 
We also show that for given set of sightlines, 
one can choose between a finer map resolution, $\sigthreed$, or a better reconstruction fidelity, \snreps.

We also derived an analytic expression that allows us to compute
\snreps\ as a function of $\sigthreed$ and signal-to-noise properties
of a given set of sightlines. This shows the dominance, on scales
$\sigthreed \gtrsim 3\,\mpc$, of shot-noise in the sampling of the
sightlines --- this argues, on these coarser scales, for obtaining
additional, fainter, sightlines over increasing \texp\ to improve
\snreps\ or reduce \sigthreed.
 
These findings motivate a survey targeted at obtaining \lya\ forest
spectra from faint $\glim \approx 24$ LBGs over $\sim 1$ square degree
on existing 8-10m telescopes and spectrographs, which would enable
tomographic reconstructions with resolutions of $\sigthreed \sim
3-4\,\mpc$ over $\sim 10^6\,\mpccube$ of comoving volume.  Such a
survey would require exposure times of $4-5\,\hrs$ on VLT-VIMOS or
$2-2.5\,\hrs$ on Keck LRIS, with a total time requirement of $\sim
130\,\hrs$ (including overheads) to cover 1 square degree.  These
exposure times and magnitude limits are comparable to contemporary
high-$z$ galaxy redshift surveys
\citep[e.g.,][]{steidel:2004,lilly:2007,le-fevre:2013}, but the data
from these surveys are not suitable for tomographic reconstruction
since they are of low resolution ($R \lesssim 400$), which does not
provide adequate line-of-sight sampling for our purposes.
 
 IGM tomography is far more efficient than galaxy redshift surveys at
 mapping out $z \sim 2$ large-scale structure, particularly at low overdensities ($\rho/\langle \rho \rangle \lesssim 10$), 
 due to the $\sim
 400-500\,\mpc$ probed along the line-of-sight of \lya\ forest
 spectrum; a tomographic survey requires only a dense \emph{areal}
 sampling rather than \emph{volume} sampling that is required for
 comparable galaxy surveys.  Our fiducial 1 square degree, $\sigthreed
 \sim 3\,\mpc$ tomographic map is already a factor of $\sim 2$ better,
 resolution-wise, compared to the \citet{kitaura:2009} map of $\sim
 250000$ SDSS galaxies at $z \sim 0.1$, which had an effective
 resolution of $\sim 7\,\mpc$ over a $350^3\,\mpccube$ volume.  To
 achieve a similar galaxy number density ($n_\mathrm{gal} \approx
 0.005\,h^3\, \mathrm{Mpc}^{-3}$) at $z \sim 2$ would require
 obtaining redshifts for a \emph{volume-limited sample of $\mathcal{R}
   \approx 26$ galaxies}. 
Even with 30m-class telescopes, it would be
 very expensive to obtain such samples over cosmologically interesting
 volumes.
 
 With the wide-field spectrographs slated to become available in the
 near-future, IGM tomography will be improved in two distinct regimes:
 (a) the Subaru PFS spectrograph on the 8m Subaru telescope will allow
 sufficiently deep ($\texp \gtrsim 10\,\hrs$) exposures over $\sim 1$
 square degree fields to enable cosmography at $\sigthreed \approx
 2\,\mpc$, a mapping scale achievable through galaxy redshift surveys
 only within the immediate ($z \lesssim 0.03$) Local Universe
 \citep[e.g.,][]{courtois:2013}. ({b}) the MS-DESI spectrograph on the
 4m KPNO Mayall telescope, which would enable \lya\ tomography with
 map resolutions of $\sigthreed \sim 7\,\mpc$ over comoving volumes of
 $\approx 8\times 10^6 \,\mpccube$ per $\texp \approx 10\,\hrs$
 pointing, enabling huge volumes to be efficiently mapped.
 
 The IGM tomographic maps will open up new science possibilities at $z
 \sim 2$. In this paper we have outlined a few applications, such as
 studying galaxy environments at $z\sim 2$, searching for galaxy
 protoclusters, and measuring the topology of large-scale structure.
 However, much observational and theoretical work needs to be carried
 out before \lya\ forest tomography becomes a useful scientific tool.
 We therefore invite the community to contribute their intellectual
 energies to this exciting and potentially very fruitful new
 observational technique.

\acknowledgements{
We thank Alberto Rorai for his assistance in creating the simulated \lya\ forest skewers, and Xavier Prochaska for
helpful discussions and advice. RC and MO acknowledge support from the grant NSF-AST 1109730. 
The Wiener reconstructions discussed in this work were performed on the THEO cluster
of the Max-Planck-Institut f\"ur Astronomie at the Rechenzentrum in Garching, Germany. 
} \\

\bibliographystyle{apj}
\bibliography{lyaf_kg,apj-jour,lss_galaxies}

\end{document}